\documentclass[reprint, prb, aps, floatfix, nofootinbib, superscriptaddress]{revtex4-2}
\usepackage{placeins}
\usepackage{comment}
\usepackage{amsmath}
\usepackage{rotating}
\usepackage{graphicx}
\usepackage{chemformula}
\usepackage{mathtools}
\usepackage{amssymb}
\usepackage{tabularx}
\usepackage{color}
\usepackage{bm}
\usepackage{dcolumn}
\usepackage[colorlinks=true, linkcolor=blue, citecolor=blue, urlcolor=blue]{hyperref}
\usepackage{siunitx}

\renewcommand{\mathbf}[1]{\bm{#1}}

\newcommand{\aNPR}{\texorpdfstring{\ensuremath{\alpha\text{-}\text{Na}_2\text{Pr}\text{O}_{3}\,}}{a-Na2PrO3}}
\newcommand{\bNPR}{\texorpdfstring{\ensuremath{\beta\text{-}\text{Na}_2\text{Pr}\text{O}_{3}\,}}{b-Na2PrO3}}
\newcommand{\LiIrO}{\texorpdfstring{\ensuremath{\text{Li}_2\text{Ir}\text{O}_{3}\,}}{Li2IrO3}}
\newcommand{\NaIrO}{\texorpdfstring{\ensuremath{\text{Na}_2\text{Ir}\text{O}_{3}\,}}{Na2IrO3}}

\newcommand{\pr}{\ensuremath{\text{Pr}^{4+}}}
\newcommand{\na}{\ensuremath{\text{Na}^{+}}}
\newcommand{\ox}{\ensuremath{\text{O}^{2-}}}

\DeclareMathOperator{\sign}{sign}

\begin{document}

\title{Canted antiferromagnetism and spin-flop transition in honeycomb $\alpha$-\ch{Na2PrO3} \\ revealed by angle-dependent torque magnetometry up to 35~T}

\author{Kylie~MacFarquharson}
\affiliation{Clarendon Laboratory, University of Oxford Physics Department, Parks Road, Oxford OX1 3PU, UK}
\author{Ryutaro~Okuma}
\affiliation{Clarendon Laboratory, University of Oxford Physics Department, Parks Road, Oxford OX1 3PU, UK}
\affiliation{Institute for Solid State Physics, University of Tokyo, Kashiwa, Chiba 277-8581, Japan}
\author{Matthew~Pearce}
\altaffiliation{Present address: Durham University, Department of Physics, South Road, Durham, DH1 3LE, UK}
\affiliation{Clarendon Laboratory, University of Oxford Physics Department, Parks Road, Oxford OX1 3PU, UK}
\author{Amalia~I.~Coldea}
\affiliation{Clarendon Laboratory, University of Oxford Physics Department, Parks Road, Oxford OX1 3PU, UK}
\author{Alix~McCollam}
\altaffiliation{Present address: School of Physics, University College Cork, Cork, Ireland}
\affiliation{High Field Magnet Laboratory (HFML-EMFL), Radboud University, Nijmegen 6525 ED, The Netherlands}
\author{Roos~Leenen}
\affiliation{High Field Magnet Laboratory (HFML-EMFL), Radboud University, Nijmegen 6525 ED, The Netherlands}
\author{Uli~Zeitler}
\affiliation{High Field Magnet Laboratory (HFML-EMFL), Radboud University, Nijmegen 6525 ED, The Netherlands}
\author{Yurii~Skourski}
\affiliation{Hochfeld-Magnetlabor Dresden (HLD-EMFL), Helmholtz-Zentrum Dresden-Rossendorf (HZDR), Dresden 01328, Germany}
\author{Radu~Coldea}
\affiliation{Clarendon Laboratory, University of Oxford Physics Department, Parks Road, Oxford OX1 3PU, UK}

\begin{abstract}
We report angle-dependent piezocantilever torque measurements on untwinned single crystals of the honeycomb magnet $\alpha$-\ch{Na2PrO3}, proposed as a candidate to display unconventional magnetic behavior stabilized by anisotropic exchange interactions between spin-orbit entangled, effective pseudospin 1/2 Pr$^{4+}$ magnetic moments. Previous powder susceptibility measurements observed signatures of a transition at low temperatures to antiferromagnetic order with a weak ferromagnetic moment. From the torque angular dependence in three orthogonal crystallographic planes we infer a zero-field canted antiferromagnetic structure, with the magnetic moments almost normal to the honeycomb layers, but weakly canted such as to give rise to a spontaneous ferromagnetic moment along the in-plane monoclinic $a$-axis, with two magnetic domains related by time reversal. Sharp jumps in the angular dependence of the torque when the field projection along the $a$-axis changes sign are interpreted as field-induced switching between the two degenerate magnetic domains with oppositely aligned ferromagnetic moments. We propose a minimal spin Hamiltonian including easy-axis exchange anisotropy and symmetry-allowed Dzyaloshinskii-Moriya (DM) couplings, for which a mean-field calculation reproduces the main features of the observed angle-dependent torque. Torque measurements in fields up to 35~T applied almost normal to the layers observe a strong anomaly, which we attribute to proximity to the field-induced spin-flop transition predicted for the above Hamiltonian. We contrast the predicted magnetic phase diagrams for magnetic fields in different orthogonal crystallographic planes normal to the honeycomb layers and discuss how DM interactions convert some transitions into smooth crossovers and stabilize a phase boundary line connecting the spin-flop transition in field normal to the layers to magnetization saturation for in-plane field normal to the $a$-axis; such a phase boundary line does not exist in the absence of DM interactions.
\end{abstract}

\date{\today}
\maketitle

\section{Introduction} 

    Magnetic insulators built from threefold-coordinated, edge-sharing octahedra of heavy transition-metal or rare-earth ions have emerged as a central materials platform for realizing bond-dependent anisotropic exchange in the strong spin-orbit regime \cite{chaloupka2010kitaev}. In such compounds, the combined effect of spin-orbit coupling and crystal-field splitting can stabilize spin-orbit-entangled $j_{\rm eff}=1/2$ magnetic moments, while the edge-sharing bonding geometry could lead to highly bond-directional superexchange interactions. This could provide the microscopic setting for realizing experimentally Kitaev \cite{kitaev2006anyons} and in general, quantum spin compass models on honeycomb and related lattice geometries. Layered honeycomb iridates Na$_2$IrO$_3$ \cite{HwanChun2015} and $\alpha$-Li$_2$IrO$_3$ \cite{Williams2016}, $\alpha$-RuCl$_3$ \cite{Banerjee2017}, the three-dimensional hyperhoneycomb $\beta$- \cite{takayama2015hyperhoneycomb,biffin2014unconventional} and stripyhoneycomb $\gamma$-Li$_2$IrO$_3$ \cite{modic2014realization,biffin2014noncoplanar}, and more recently also hyperhoneycomb $\beta$-Na$_2$PrO$_3$ \cite{okuma2023noncollinear} have been intensely investigated as candidate Kitaev materials in which frustrated anisotropic exchanges stabilize unconventional ordered states and/or spin dynamics.

    A key conclusion from those studies is that the balance between Kitaev, Heisenberg, off-diagonal symmetric exchange and longer-range interactions depends very sensitively on the detailed crystalline environment and the specific characteristics of the orbitals involved in the superexchange \cite{Rousochatzakis2024}. It is therefore insightful to explore experimentally other materials that also have the key ingredients to potentially realize such interactions. The layered magnet \aNPR \cite{hinatsu2006crystal,Ramanathan2021,synthesis} has a monoclinic crystal structure (space group $C2/c$) with Pr$^{4+}$ ions occupying a layered honeycomb network of edge-sharing octahedra, which could promote bond-dependent anisotropic exchange. Recent theoretical studies \cite{motome2020materials} predicted substantial Kitaev exchanges in the limit of weak octahedral crystal field relative to the spin orbit coupling, and under certain simplifying assumptions about the nature of the superexchange processes. On the experimental side, inelastic neutron scattering measurements of crystal-field transitions \cite{Onuorah2024} found evidence that the magnetic ground state departs from the idealised $j_{\rm eff}=1/2$ expected for Pr$^{4+}$ ions in the weak octahedral cubic crystal field limit. Magnetic susceptibility and $\mu$SR measurements on powder samples found evidence for a transition below $T_{\rm N}=4.6$~K to antiferromagnetic long-range order with a weak ferromagnetic moment, and powder inelastic neutron scattering measurements observed a gapped spectrum and suggested dominant easy-axis XXZ exchange interactions between nearest neighbors \cite{daum2021collective,Onuorah2024}.     

    A very sensitive probe of the magnetic anisotropy is provided by measurements of the magnetic torque as a function of the applied magnetic field orientation relative to the crystallographic axes. We have recently succeeded in synthesizing single crystals of \aNPR \cite{synthesis} and here we report angle-dependent magnetic torque measurements on untwinned samples in static magnetic fields up to 35~T. We infer a canted N\'{e}el order with the main antiferromagnetic components normal to the honeycomb layers and a small spontaneous ferromagnetic component along the in-plane $a$-direction. We propose a minimal spin Hamiltonian with nearest-neighbor XXZ exchange on the honeycomb lattice accompanied by symmetry-allowed DM interactions and find good agreement between the observed angular dependence of the torque in three orthogonal crystallographic planes and mean-field calculations for this Hamiltonian. Clear jumps observed in the angle-dependent torque data are attributed to field-induced switching between the two time-reversed magnetic domains with oppositely aligned spontaneous magnetization. Torque anomalies observed in much higher fields applied almost normal to the honeycomb layers are interpreted as evidence for a field-induced spin-flop transition.  

    The remainder of this paper is organized as follows. Sec.~\ref{S:exp} gives details of the experimental setup used for the magnetic torque measurements. The following Sec.~\ref{S:torque16T} reports the torque angular dependence for field applied in three orthogonal crystallographic planes, from which we infer a model for the low-field magnetic structure in Sec.~\ref{aNPR:sec:magnetic}, described in terms of magnetic basis vectors compatible with the crystal structure symmetry in Sec.~\ref{aNPR:sec:mag_sym}. A minimal spin Hamiltonian with easy-axis XXZ exchange anisotropy accompanied by symmetry-allowed DM interactions (with an effective DM vector along the two-fold $b$-axis) is proposed in Sec.~\ref{aNPR:sec:mean-field}, which also shows mean-field calculations compared with the measured torque data. The calculated magnetic phase diagrams for the minimal model for field applied in a plane normal to the honeycomb layers and containing the effective DM vector or normal to it are very different and the important role played by the DM interactions in changing the nature of the phases and/or phase transitions is discussed. Sec.~\ref{aNPR:sec:high-field} presents torque measurements in fields up to 35~T that observe clear anomalies, proposed to be associated with a field-induced transition to an intermediate-field spin-flop state, as expected for the minimal Hamiltonian. The results are compared to other honeycomb Kitaev candidate materials in Sec.~\ref{aNPR:sec:discussion}. Finally, Sec.~\ref{aNPR:sec:conclusion} summarizes the main results and conclusions. For completeness, Appendix~\ref{aNPR:sec:xrays} presents structural characterization of the samples using X-ray diffraction, confirming that the samples were untwinned, single-grain with no evidence for structural defects such as layer stacking faults or atomic site mixing. Appendix~\ref{aNPR:sec:misalignment} considers the sensitivity of the torque angular dependence to a small misalignment of the rotation axis. Appendix~\ref{aNPR:sec:calib} presents technical details of the thermometer calibration for the measurements in magnetic fields up to 35~T. 

\section{Experimental Details}\label{S:exp}
	Crystals of \aNPR were synthesized as explained in Ref.~\cite{synthesis}. The samples used for the torque measurements were thin plates of $\sim$80~$\mu$m diameter. X-ray diffraction measurements (for details see Appendix~\ref{aNPR:sec:xrays}) confirmed they were untwinned, single-grain crystals, with the nominal monoclinic $C2/c$ crystal structure. Torque measurements were performed using Seiko PRC400 piezocantilevers mounted on a rotating platform inside a 16~T Quantum Design PPMS system with a base temperature of 2~K. Additional torque measurements in higher fields up to 35~T and temperatures down to 3~K were performed using the same cantilever platforms at the High Field Magnet Laboratory in Nijmegen. The samples were synthesized and stored in an inert gas atmosphere and were exposed to ambient air conditions for only short amounts of time, during the X-ray screening (typically under one hour, covered in a thin layer of vacuum grease) and during mounting onto the torque cantilever platform (again using a small amount of vacuum grease), after which they were immediately cooled under a $^4$He exchange gas atmosphere. 

    In an applied field $\mathbf{H}$, if the magnetic moment of the sample $\mathbf{m}$ is not parallel to the applied field, the sample will be subject to a torque $\mathbf{\tau}=\mu_0\mathbf{m}\times\mathbf{H}$. The torque component $\tau_{u}=\mathbf{\tau}\cdot\mathbf{u}$ along the cantilever bending axis $\mathbf{u}$ will bend the cantilever, changing the resistance of the piezocrystals at its base. This is detected experimentally using a Wheatstone bridge via a voltage change $\Delta V$, with a linear dependence between $\Delta V$ and $\tau_{u}$ for small cantilever angular deviations, expected to be the case for all measurements reported here. All plotted angular torque curves $\tau_u(\theta)$ have a vertical offset subtracted off to ensure that the average over a 180$^{\circ}$ angular range is zero, as expected for magnetic torque. The torque $\tau_{u}$ is plotted in arbitrary units (arb. units), but the absolute sign is meaningful. It is determined from the observed angular dependence of the torque in zero applied field, when the very small, but finite sample weight gives a characteristic torque signal (this non-magnetic torque signal is negligible relative to the observed magnetic torque so it was not subtracted off from the raw torque data). The absolute sign of the torque was also confirmed by measurements using the same experimental setup on single crystal samples of magnetic materials with well-characterized magnetic anisotropy.

\begin{figure}[tb]
	\includegraphics[width=0.5\textwidth]{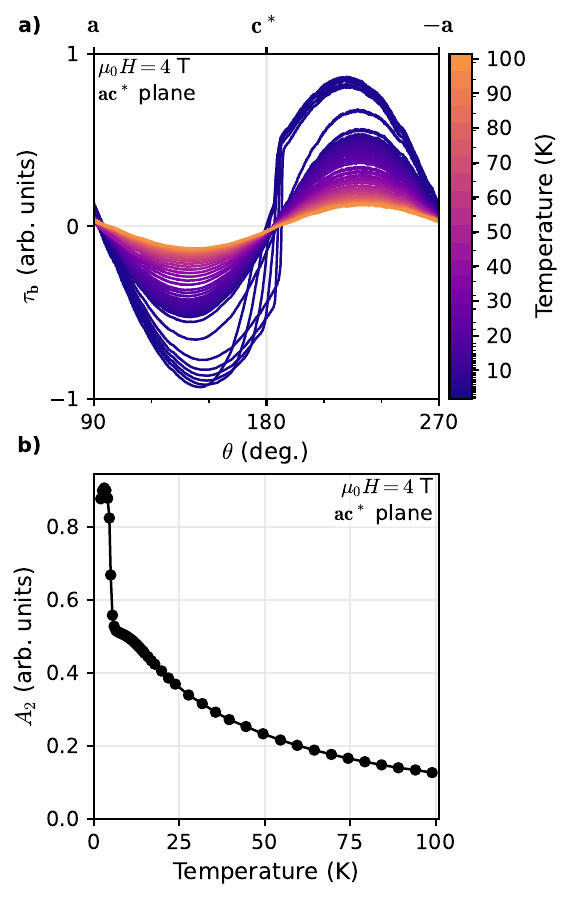}\centering
	\caption[Angular dependence of the magnetic torque as a function of temperature in the $ac$ plane.]{(a) Angular dependence of the magnetic torque as a function of temperature (color-coded) between 2 and 100~K, in a field of 4~T in the $ac^*$ plane, with the sample platform rotation axis $\mathbf{u}\parallel \mathbf{b}$. Top labels indicate when the applied field is along reference crystallographic axes. (b) Temperature-dependence of the amplitude of a sinusoidal curve fitted to the torque data in (a) as per eq.~(\ref{aNPR:eqn:2theta}).}\label{aNPR:fig:torque:temperature}
\end{figure}

    	\begin{figure}[tb]
        \centering
		\includegraphics{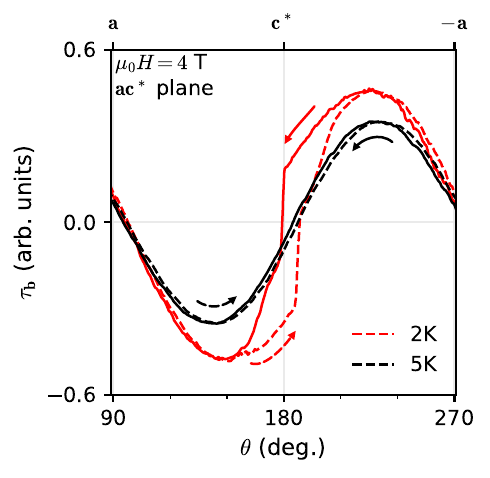}
		\caption[Angular dependence of torque in the $ac^*$ plane showing both rotation directions.]{Contrasting the angular dependence of the torque for field in the $ac^*$ plane, below (red) and above (black) the magnetic ordering transition temperature $T_{\rm N}$, for clockwise (dashed) and anticlockwise senses (solid lines) of rotation, also indicated by the overlaid arrows. Note the sharp jumps and hysteresis observed at low temperatures are absent above $T_{\rm N}$. Top labels indicate when the applied field is along reference crystallographic axes.}
		\label{aNPR:fig:torque:hysteresis}
	\end{figure}
    
    To probe the angular dependence of the torque for magnetic field in three orthogonal crystallographic planes, samples were mounted in three distinct orientations relative to the cantilever platform. The cantilever bending axis $\mathbf{u}$ was normal to the (fixed) applied field $\mathbf{H}$ and parallel to cantilever platform rotation axis. 
    Measurements were performed by rotating the cantilever platform angle $\theta$ over an angular range of 180 or 360$^{\circ}$, in both the clockwise and the anticlockwise senses and unless explicitly stated otherwise all data reported here are for the anticlockwise sense. The torque component along the rotation axis is related to the angular derivative of the free energy $\mathcal{F}$ via $\tau_u=-\frac{\partial \mathcal{F}}{\partial \theta}$, where the minus sign comes from the fact that in the experiment the sample is rotated around axis $\mathbf{u}$ by an angle $\theta$ and the field $\mathbf{H}$ is fixed, rather than the sample being fixed and the field being rotated by angle $\theta$. An angle $\theta_0$ corresponds to a stable equilibrium if the torque $\tau_u$ is zero with a negative gradient, i.e. for small deviations away from $\theta_0$ the torque has a restoring form $\tau_u(\theta)=-\kappa(\theta-\theta_0)$ with $\kappa>0$. Throughout we specify the applied field direction with reference to the set of mutually orthogonal axes $\mathbf{a}$, $\mathbf{b}$ and $\mathbf{c^*} \parallel \mathbf{a} \times \mathbf{b}$, where $\mathbf{a}$, $\mathbf{b}$, and $\mathbf{c}$ are the monoclinic crystallographic axes.         
    
    \section{Angle-dependent torque up to 4~T}\label{S:torque16T}
    \subsection{Torque for field in the $ac^*$ plane}
    
    An overview of the angular dependence of the torque for field in the $ac^*$ plane as a function of temperature is shown in Fig.~\ref{aNPR:fig:torque:temperature}(a). At high temperatures, the torque has a two-fold sinusoidal dependence as expected for low fields in the paramagnetic regime~\cite{xiao2006angular, zhang2015spin}. The stable equilibrium (zero torque with negative gradient) corresponds to field along $a$ and unstable equilibrium (zero torque with positive gradient) for field along $c^*$, which gives the order of the magnetic susceptibilities in the paramagnetic phase as $ \chi_a>\chi_{c^*}$. The angular dependence is \begin{eqnarray}
		\tau_b(\theta) &=& A_2 \sin 2(\theta-\theta_{c^*}) \label{aNPR:eqn:2theta}\\
		&=&\frac{1}{2} \mu_0H^2 \left(\chi_{a} - \chi_{c^*}\right)  \sin 2(\theta-\theta_{c^*}) 
        \label{aNPR:eqn:B2}\nonumber
	\end{eqnarray}
	where $\theta_{c^*}$ is the rotation angle when the field is along $c^*$. 
	
    \begin{figure*}[tb]
		\includegraphics{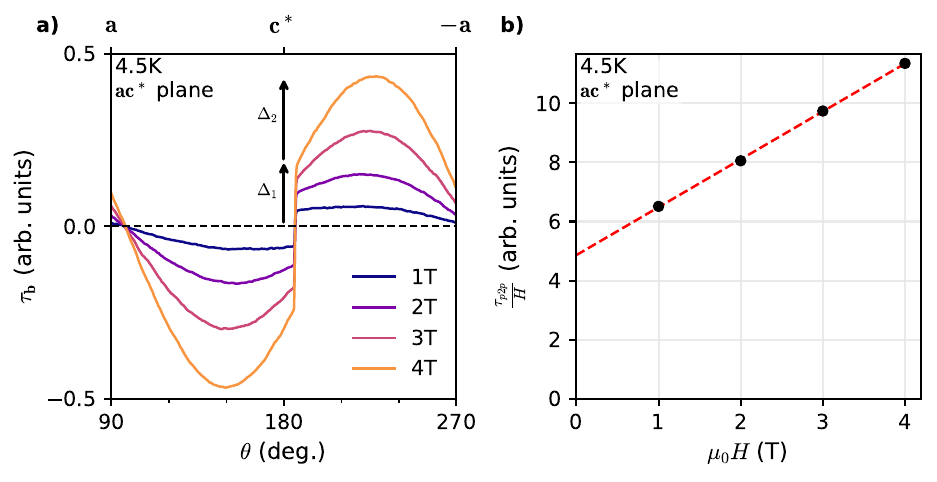}
		\caption[Angular and field dependence of magnetic torque of \aNPR in the  $ac$ plane.]{(a) Angular dependence of the torque in the $ac^*$ plane at 4.5~K just below $T_{\rm N}$, as a function of field magnitude (color-coded). Top labels indicate when the field direction is along reference crystallographic axes. (b) Peak-to-peak torque bandwidth $\tau_{\rm{p2p}}$ divided by field strength $H$ as a function of field, fitted to a linear behavior (dashed line).} 
		\label{aNPR:fig:torque:fields}
	\end{figure*}
    
	The torque data were fitted to the sinusoidal form in eq.~(\ref{aNPR:eqn:2theta}) and the resulting amplitude $A_2$ as a function of temperature is plotted in Fig.~\ref{aNPR:fig:torque:temperature}(b). The amplitude increases upon cooling, has a broad hump around 10~K, and a clear sharp upturn upon crossing $T_{\rm N}=4.6$~K. It is below this temperature that previous powder susceptibility measurements \cite{daum2021collective} observed a splitting between ZFC (zero-field cooled) and FC susceptibilities, attributed to the onset of long-range magnetic order.

    At temperatures below $T_{\rm N}$ the torque profile no longer has a simple sinusoidal shape. This can be seen most clearly in Fig.~\ref{aNPR:fig:torque:hysteresis}, which shows a comparison of the angular dependence between just above (5~K, black curves) and well below $T_{\rm N}$ (2~K, red curves). Both senses of rotation are shown, clockwise ($\theta$ increasing) as dashed lines, and anticlockwise ($\theta$ decreasing) as solid lines. Above $T_{\rm N}$ the torque is sinusoidal, with the unstable equilibrium for field near $c^*$ and with the clockwise and anticlockwise curves approximately the same, up to a very small ($\sim 1^\circ$) offset in $\theta$ caused by backlash in the rotator probe. However, distinct features appear below $T_{\rm N}$. There is significant hysteresis between the clockwise and anticlockwise senses, and a sharp jump in torque is observed as the field crosses the $c^*$ axis when the projection of the applied field onto the $a$ axis changes sign. The angular width of the hysteresis is larger for lower temperatures and the precise magnitude is sample dependent. We attribute the sharp jump and hysteresis to the onset of magnetic order. The (positive) area enclosed by the hysteresis cycle, $-\oint \tau_u d\theta$, is the net work done in one full rotation cycle by the rotating motor of the platform against the magnetic domain pinning forces in the sample.

    \begin{figure}[tb]
		\includegraphics[width=0.5\textwidth]{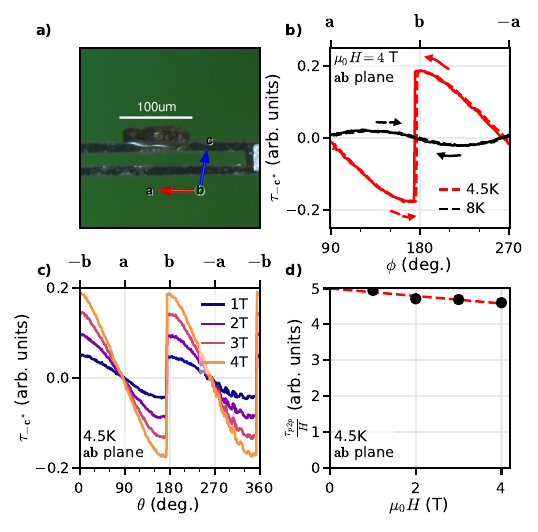}\centering
		\caption[Angular dependence of magnetic torque in $ab$ plane.]{(a) Photograph of the mounted sample for torque measurements for field in the $ab$ plane. The sample is visible just below the scale bar in the center of the image, attached to the side of the horizontal two-prong cantilever. (b) Angular dependence of the torque in an applied field of 4~T, just below $T_{\rm N}$ (red, 4.5~K) and above (8~K, black). Dashed/solid traces correspond to the clockwise/anticlockwise sense of rotation, also indicated by overlaid arrows. (c) Torque angular dependence as a function of field magnitude at 4.5~K. Top labels in (b) and (c) indicate when the field is along reference crystallographic axes. (d) Field dependence of the peak-to-peak torque bandwidth divided by field, fitted to a linear behavior (dashed line).}
		\label{aNPR:fig:torque:ab}
	\end{figure}
    
	Fig.~\ref{aNPR:fig:torque:fields}(a) shows the angular dependence of the torque as a function of increasing field at 4.5~K. This temperature was chosen as it is just below $T_{\rm N}$, where the torque jumps are particularly sharp and the hysteresis is not too large such that one can still find a relatively simple parameterization of the torque shape. There are two contributions to the overall bandwidth of the torque curves, the half-height of the torque jump $\Delta_1$ and the remaining half-height of the bandwidth $\Delta_2$, see Fig.~\ref{aNPR:fig:torque:fields}(a). We find empirically that their field scaling is consistent with $\Delta_1 \sim H$ and $\Delta_2 \sim H^2$ (see Fig.~\ref{aNPR:fig:torque:fields}(b)), suggesting that they have different origins.    
       
    \subsection{Torque for field in the $ab$ plane}
    
\begin{figure}
		\includegraphics[width=0.47\textwidth]{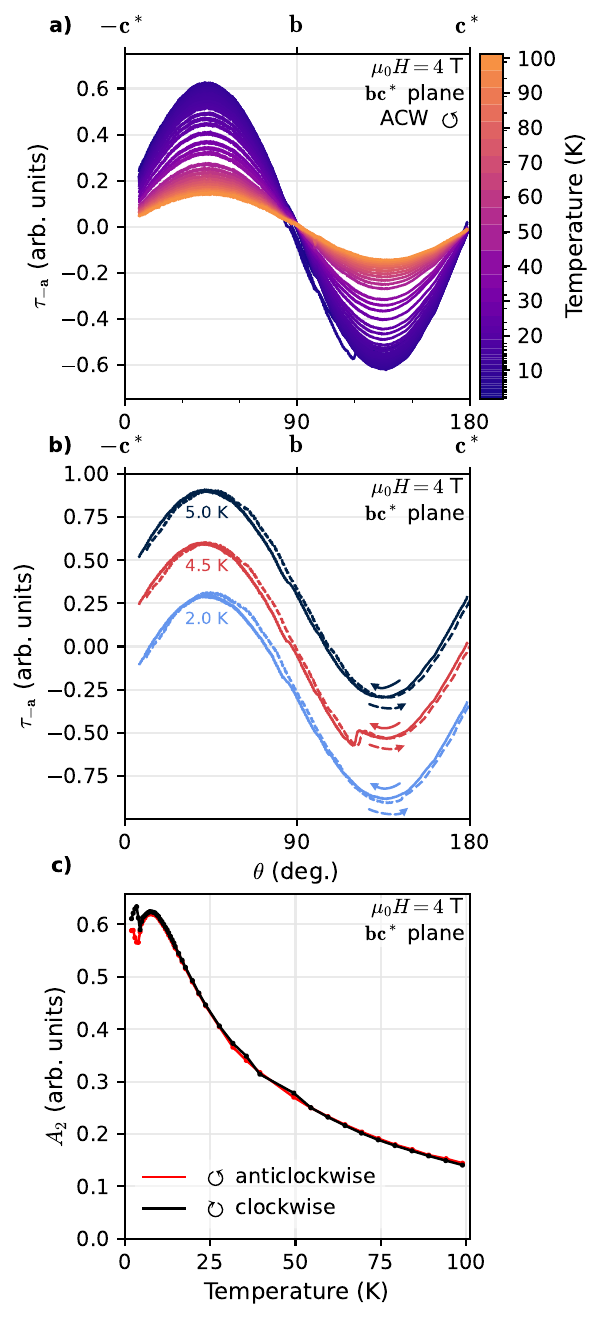}
		\caption[Angular dependence of magnetic torque in $bc^*$ plane.]{(a) Torque angular dependence for field in the $bc^*$ plane as a function of temperature (color-coded) between 2 and 100~K. Top labels indicate when the field is along reference crystallographic axes. (b) Comparison of CW (dashed) and ACW (solid) curves above $T_{\rm N} $ at 5~K (upper traces), at 4.5~K just below $T_{\rm N}$ (middle traces) and at 2~K (lower traces), offset vertically for clarity by $0.3, 0$ and $-0.3$, respectively. The small jump anomaly observed in the middle traces is attributed to domain switching in the presence of weak domain pinning effects and a small crystal misalignment as discussed in the text and quantitatively parameterized in Appendix~\ref{aNPR:sec:misalignment}. (c) Temperature dependence of the fitted torque amplitude $A_2$ with black/red curves corresponding to clockwise/anticlockwise senses of rotation.}
		\label{aNPR:fig:torque:bc}
	\end{figure}

	To further explore the magnetic anisotropy, the angular dependence of the torque was measured also for field in the other two orthogonal crystallographic planes $ab$ and $bc^*$. The air sensitivity of the samples, and the need for them to be exposed to atmosphere during mounting meant that it was not possible to measure the same sample in all three orientations and be confident that it survived multiple re-mountings unaffected. The same sample was used for the data in Figs.~\ref{aNPR:fig:torque:temperature} ($ac^*$ plane) and ~\ref{aNPR:fig:torque:ab} ($ab$ plane), but a different sample in Fig.~\ref{aNPR:fig:torque:bc} ($bc^*$ plane). The torque units between different samples are not directly comparable.  
    
	As the samples are platelets in the $ab$ plane, for measurements in the $ac^*$ and $bc^*$ planes the sample plate is mounted directly on top of the flat cantilever plane [see Fig.~\ref{aNPR:fig:xrays}(c)], whereas for measurements in the $ab$ plane the mounting is much more challenging as the sample plate needs to be normal to the cantilever plane and the sample is now attached to the side of the very thin cantilever, held in place by a small amount of chilled vacuum grease, see Fig.~\ref{aNPR:fig:torque:ab}(a).

    Fig.~\ref{aNPR:fig:torque:ab}(b) shows the measured torque angular dependence for field in the $ab$ plane. The high temperature data (8~K, black) shows a very small, but finite 2-fold periodic torque signal with no observable hysteresis. In this regime, the stable equilibrium occurs for field along $b$-axis and unstable equilibrium for field along $a$, which implies $\chi_b>\chi_a$. We attribute this anisotropy to a small difference in the $g$-factor along the two directions $g_b > g_a$, symmetry-allowed by the monoclinic crystal structure. Below $T_{\rm N}$ (red), the torque signal is much bigger, with the stable equilibrium now shifted to field along $a$ and $\chi_a>\chi_b$, with a clear jump in torque when the projection of the field along the $a$-axis changes sign. 
	
	Fig.~\ref{aNPR:fig:torque:ab}(c) shows the angular dependence as a function of field magnitude (color-coded). The same features are seen at all measured fields, with a larger torque amplitude at higher fields. The peak-to-peak torque bandwidth scaled by field $\dfrac{\tau_{p2p}}{H}$ is plotted in Fig.~\ref{aNPR:fig:torque:ab}(d). It is approximately constant, demonstrating that the size of the jump is almost linear in magnetic field, as was seen also in the $ac^*$ plane. A small negative gradient in the linear fit is attributed to the small difference in $g$-factors between the $a$ and $b$ directions, as deduced from Fig.~\ref{aNPR:fig:torque:ab}(b) (black curves).
	
    \subsection{Torque for field in the $bc^*$ plane}
    
    Finally, the torque angular dependence for field in the $bc^*$ plane is shown in Fig.~\ref{aNPR:fig:torque:bc}(a). The torque shape is a two-fold periodic, sinusoidal curve with the stable and unstable equilibrium positions for field along $b$ and $c^*$ axes, respectively, so $\chi_b>\chi_{c^*}$ at all probed temperatures. The torque angular dependence does not show very prominent jumps below $T_N$ (in contrast to measurements in the $ac^*$ and $ab$ planes), however a very small step-like anomaly is detected just below $T_{\rm N}$ as illustrated in Fig.~\ref{aNPR:fig:torque:bc}(b) (middle traces). 
    We attribute this to a slight misalignment of the sample rotation axis relative to the nominal $a$-axis. In this case the magnetic field will have a very small, but finite component along the $a$-axis, with $H_a>0$ for half the rotation plane and $H_a<0$ for the other half, so domain switching and a small jump in torque is still expected when the rotation angle is such that the small but finite $H_a$ changes sign, for more details see Appendix~\ref{aNPR:sec:misalignment}. The small jump anomaly is not present in the paramagnetic phase (5~K, upper traces), and is also suppressed at much lower temperatures (2~K, lower traces). We propose that the reason for the latter is that the very small applied field component onto the $a$-axis is not strong enough to overcome the increased domain pinning effects at those lower temperatures.    
    
    As a consequence of the small hysteresis effect observed just below $T_{\rm N}$, the amplitudes of the CW and ACW torque curves when fitted to a sinusoidal form as in eq.~(1) differ slightly below $T_{\rm N}$ as illustrated in Fig.~\ref{aNPR:fig:torque:bc}(c). As the observed torque anomalies are very small relative to the overall torque bandwidth in this plane, we expect that for the idealized case of a perfectly aligned sample there would be no jump anomalies (and no hysteresis) in torque, so no domain switching. We therefore conclude that the $bc^*$ plane is (within our experimental accuracy) normal to the direction of the spontaneous zero-field ferromagnetic moment, which we henceforth assume is along the $a$-axis. 
    
\section{Magnetic Structure}
\label{aNPR:sec:magnetic}

	These results can be naturally explained by a canted magnetic structure, with the dominant antiferromagnetic components along the $c^*$-axis and a small, spontaneous ferromagnetic moment along the $a$-axis. In this case one expects two magnetic domains related to one another by time-reversal. An applied magnetic field with a non-zero component along the $a$-axis then selects a single magnetic domain. Changing the sign of $H_a$ by passing through zero induces a switch between the two magnetic domains, and this is accompanied by a jump in torque, as we show below. 

    Consider that the two time-reversed magnetic domains have a spontaneous magnetic moment of magnitude $m$ along the $\pm a$ axis. At leading order in the applied field strength, the torque will be dominated by the contribution from the spontaneous magnetic moment of magnitude $m$, as if it was constant and pinned to the $\pm a$ axes, i.e. one can neglect the torque contribution from the small field-induced magnetization. In this case, the magnetic energy is 
	\begin{eqnarray}
        E	&=& -\mu_0\bm{m} \cdot \bm{H} \nonumber \\
        &=& \mp \mu_0m H \cos(\theta-\theta_a)=\pm \mu_0mH\sin(\theta-\theta_b)
        \label{aNPR:eqn:energy}
	\end{eqnarray}
	where for concreteness we consider field applied in the $ab$ plane as for the experiments for which the torque data is plotted in Fig.~\ref{aNPR:fig:torque:ab}(c), where $\theta_{a,b}$ correspond to the rotation angle when the field is along the $a$($b$) axis. For $\sin(\theta-\theta_b) > 0$ the minimum energy corresponds to the lower sign, i.e. the ``$-$" domain with spontaneous ferromagnetic moment along $-a$. For $\sin(\theta-\theta_b) < 0$ the minimum energy corresponds instead to the ``$+$" domain. At a general $\theta$ the minimum energy domain then has the energy 
    \begin{equation}
		E(\theta)	=- \mu_0mH \sin (\theta-\theta_b) \sign (\sin (\theta-\theta_b)).\nonumber		
	\end{equation}
    The corresponding torque along the rotation axis \mbox{$\mathbf{u} \parallel -\mathbf{c^*}$} is obtained from the energy derivative as  
	\begin{equation}
    \tau_{-c^*} = -\dfrac{\partial E}{\partial \theta}= \mu_0mH \cos(\theta-\theta_b)\sign (\sin (\theta-\theta_b)), \nonumber
	\end{equation}
	so the torque is linear in the field $H$, with a $\cos(\theta-\theta_b)$ angular dependence, with a jump from $-\mu_0mH$ to $+\mu_0mH$ upon increasing $\theta$ through $\theta_b$. This can naturally explain the observed angular and field dependence of the torque in Fig.~\ref{aNPR:fig:torque:ab}(c). 
    
    \section{Magnetic Symmetry Analysis}
    \label{aNPR:sec:mag_sym}
    There are four Pr ions in the structural primitive cell, split equally between two distinct sublattices, \pr $(1)$  and $(2)$. Since this number of sites in the primitive cell can already accommodate both ferromagnetic and antiferromagnetic order patterns, we search for a minimal magnetic structure model within this cell, i.e. with magnetic propagation vector $\mathbf{q} = 0$. The irreducible representations (irreps) of symmetry-allowed $\mathbf{q} = 0$ magnetic structures and their magnetic basis vectors are tabulated in Table~\ref{aNPR:tab:irreps}. 
    \begin{table}[b]
    \caption{\label{aNPR:tab:irreps}Irreducible representations (irreps), magnetic basis vectors and magnetic space groups for $\mathbf{q}\! =\! \bf{0}$ magnetic structures obtained using \textsc{isodistort} \cite{isotropy}. $x$, $y$ and $z$ denote Cartesian axes along $\mathbf{a}$, $\mathbf{b}$ and $\mathbf{c^*}$, respectively. Last column gives the unit cell vectors of the magnetic space group in terms of the structural unit cell vectors.}
    \begin{ruledtabular}
    \begin{tabular}{cccc}
    \textrm{Irrep}&
    \textrm{Magnetic}&
    \textrm{Magnetic}&
    \textrm{Magnetic}\\
    \textrm{}&
    \textrm{Basis Vectors}&
    \textrm{Space Group}&
    \textrm{Unit Cell}\\
    \colrule
    \noalign{\vskip 1pt}
    	$m\Gamma_1^- $ & $A_y$,  $A_y'$ & $C2/c'$ &($-\mathbf{a},\mathbf{b},-\mathbf{c}$) \\
    	$m\Gamma_1^+ $ & $F_y$, $F_y'$ & $C2/c$ &($-\mathbf{a},\mathbf{b},-\mathbf{c}$)\\
    	$m\Gamma_2^- $ & $A_x$, $A_z$, $A_x'$, $A_z'$ & $C2'/c$ &($-\mathbf{a},\mathbf{b},-\mathbf{c}$) \\
    	$m\Gamma_2^+ $ & $F_x$, $F_z$,$F_x'$, $F_z'$ & $C2'/c'$ &($\mathbf{a},\mathbf{b},\mathbf{c}$)\\
    \end{tabular}
    \end{ruledtabular}
    \label{tab:MSG}
    \end{table}
    $F(A)$ basis vectors correspond to spins ferromagnetically (antiferromagnetically) aligned between the sites of the same sublattice in adjacent layers stacked along $c$. Unprimed and primed basis vectors refer to the \pr $(1)$  and $(2)$ sublattices, respectively. $x$, $y$ and $z$ are Cartesian axes along the $\mathbf{a}$, $\mathbf{b}$ and $\mathbf{c^*}$ directions, respectively.
    
    \begin{figure}[bt]
		\includegraphics{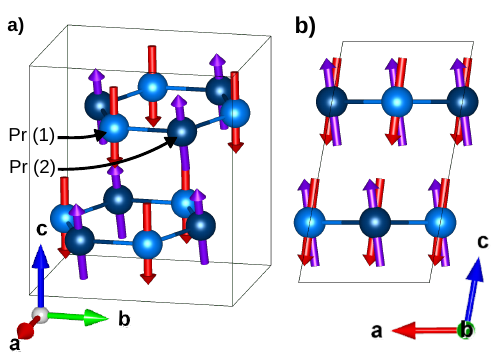}\centering
		\caption[Proposed zero-field magnetic structure of \aNPR in the monoclinic setting.]{ Proposed zero-field canted antiferromagnetic (canted AFM) structure of \aNPR. Light/dark blue spheres and red/purple arrows indicate \pr ions and spins on sublattices (1) and (2), respectively. Note that spins on the same sublattice in adjacent layers are parallel. The thin solid lines indicate the outline of the monoclinic structural unit cell, with the triad of arrows indicating the monoclinic unit cell vectors. (a) Perspective view and (b) projection onto the $ac$ plane. Spins are confined to the $ac$ plane, with the dominant components antiferromagnetically aligned along $c^*$, with a small canting (exaggerated in the figure for clarity) to give a net ferromagnetic moment along the $a$-axis. The plotted magnetic structure corresponds to the magnetic domain with the spontaneous ferromagnetic moment along the $+a$ axis. The degenerate magnetic domain with spontaneous ferromagnetic moment along the $-a$ axis is obtained by applying time reversal (reversing the orientation of all spins) or equivalently a $2_y$ two-fold rotation that passes through each Pr site.}   
		\label{aNPR:fig:magnetic}
	\end{figure}
    
	Previously reported calorimetry data \cite{daum2021collective} found that entropy was continuous through the magnetic transition upon cooling through $T_{\rm N}$. This is consistent with a continuous (second-order) transition, which in turn suggests that the magnetic order just below $T_{\rm N}$ should be described by a single irrep of the high-temperature paramagnetic space group. A ferromagnetic moment along the $a$ axis is achieved by a basis vector combination $(F_x, F'_x)$, which has spin components along the $a$-axis parallel between adjacent layers, parallel (and we assume also of equal magnitude) between the two sublattices. In order to describe the dominant antiferromagnetism along the $c^*$ axis, an additional basis vector combination of the form $(F_z, - F'_z)$ is needed, where the $-$ sign indicates that the $z$ spin components of the two Pr sites (1) and (2) are antiparallel but equal in magnitude, with parallel spins on sites of the same sublattice in adjacent layers (this antiferromagnetic order pattern is labeled N\'{e}el-I in the notation of Ref.~\cite{Onuorah2024} ). The complete magnetic structure is illustrated in Fig.~\ref{aNPR:fig:magnetic}. Note that all four contributing basis vectors are part of the same irrep $m\Gamma^+_2$, which is consistent with a continuous transition at $T_{\rm N}$ as suggested by the calorimetry data. 
    
    We note that for the canted AFM model with equal moment magnitudes at the two sites, there is no symmetry-imposed requirement that the ferromagnetic moment is constrained to be precisely along $a$ and antiferromagnetic components along $c^*$. In principle, any two orthogonal axes in the $ac^*$ plane could accommodate ferromagnetic and antiferromagnetic components in a structure described also by a (different) combination of the above same four basis vectors within the same $m\Gamma_2^+$ irrep. We have selected the particular basis vector combination $(F_x, F'_x)$ with $(F_z, - F'_z)$ because this is most consistent with the torque data, i.e. the jumps in torque occur when the field is within a few degrees of being perpendicular to the $a$ axis, which is within the accuracy of our visual alignment of the crystal on the cantilever platform. 
	
	For completeness, we note that an alternative magnetic structure to explain ferromagnetism for $a$-axis components and  antiferromagnetism for $c^*$-axis components is obtained by the basis vector combination $(F_x,F'_x)$ with $(A_z, A'_z)$, which has antiparallel $z$-axis spins between sites of the same sublattice in adjacent layers. This, however, requires mixing magnetic basis vectors from different irreps $m\Gamma^{+}_2$ and $m\Gamma^{-}_2$. A single continuous transition from paramagnetic to magnetic order at $T_{\rm N}$ directly into such a mixed-irrep state is in general not expected from Landau theory, making this scenario less consistent with the calorimetry data, which are consistent with a continuous transition.
      
	As the two \pr sites are not symmetry-equivalent, there is no symmetry constraint that the moments on the two sites are the same. A ferrimagnetic order, where these values are different, would naturally have a ferromagnetic and antiferromagnetic component. These components would however be along the same direction. For example, consider the basis vector combination $(F_x, - \alpha F'_x)$ with $0<\alpha<1$. This has a ferromagnetic net moment (per site) of $\mathbf{\hat{x}}(1 - \alpha)/2$ and an antiferromagnetic staggered moment (per site) of $\mathbf{\hat{x}}(1 + \alpha)/2$, so both components are collinear along $x$. We have calculated the torque angular dependence for candidate Hamiltonians with the above zero-field ferrimagnetic ground state, the presence of torque jumps for field in the $ac^*$ and $ab$ planes but not in the $bc^*$ plane is reproduced, however the calculated angular dependence of the torque for field in the $ac^*$ plane (not shown) cannot reproduce the additional, large sinusoidal component beyond the jump (see Fig.~\ref{aNPR:fig:torque:fields}(a)), so we discount the ferrimagnetic model. We show in the subsequent section that all features of the torque data can be naturally explained by a minimal Hamiltonian that has the zero field ground state the canted AFM structure in Fig.~\ref{aNPR:fig:magnetic}.
    
	\begin{figure}[htb]
		\includegraphics[width=0.3\textwidth]{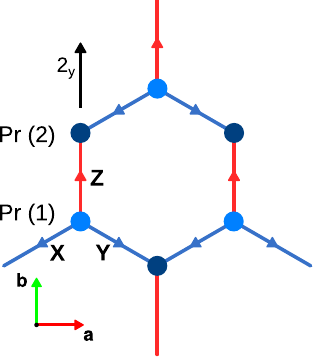}\centering
		\caption[Schematic diagram of the Pr honeycomb layer at $z=3/4$]{
        Schematic diagram of the Pr honeycomb layer at $z=3/4$. Different colors indicate symmetry-distinct sites or bonds. Arrows on bonds indicate the order of the two bond ends when expressing the bond energy in terms of a superexchange matrix, i.e. $\mathcal{H}_{\rm Z}=\sum_{\alpha,\beta\in\{x,y,z\}}S^\alpha_1\mathcal{J}^{\rm Z}_{\alpha\beta}S_2^\beta$ for a Z-bond with the exchange matrix in eq.~(\ref{aNPR:eqn:Zbond}) and $S=1/2$.}
		\label{aNPR:fig:bonds}
	\end{figure}
    
    \section{Mean-field calculations for a minimal Hamiltonian}
    \label{aNPR:sec:mean-field}

    \subsection{Minimal XXZ$+D$ Hamiltonian}
	Here we propose a minimal anisotropic exchange Hamiltonian that can stabilize the magnetic structure in Fig.~\ref{aNPR:fig:magnetic}. As illustrated in Fig.~\ref{aNPR:fig:bonds}, each honeycomb layer has two types of nearest-neighbor bonds: Z-bonds (red) along the $b$ axis, and zigzag X and Y bonds (blue), where the latter two are symmetry-related by the $2_y$ two-fold rotation that passes through each Pr site. The Z-bond has point group 2 with the unique axis along $y$, which constrains the exchange matrix to the form 
    \begin{equation}        
        \mathcal{J}^{\rm Z}=
        \begin{pmatrix}
        J'_x & 0 & \Lambda_y'-D_y' \\
        0 & J'_y & 0 \\
        \Lambda_y'+D_y' & 0 & J'_z
        \end{pmatrix},
        \label{aNPR:eqn:Zbond}
    \end{equation}
	where $\Lambda_y'$ parametrizes a symmetric off-diagonal exchange $S_1^zS_2^x+S_1^xS_2^z$ and $D_y'$ parametrizes a Dzyaloshinskii–Moriya (DM) interaction $S_1^zS_2^x-S_1^xS_2^z$. The zigzag bonds have point group 1, so the exchange matrix is completely unconstrained, with the general form for an X bond 
		\begin{equation}	
		\mathcal{J}^{\rm X}= \begin{pmatrix}
			J''_x & \Lambda''_z+D''_z & \Lambda''_y-D''_y \\
			\Lambda''_z-D''_z & J''_y & \Lambda''_x+D''_x \\
			\Lambda''_y+D''_y & \Lambda''_x-D''_x & J''_z 
		\end{pmatrix}.
        \label{aNPR:eqn:Xbond}
   		\end{equation}
	The exchange matrix for a Y bond is obtained from the above via a $2_y$ rotation as 	
	    \begin{equation}	
		\mathcal{J}^{\rm Y}= \begin{pmatrix}
			J''_x & -\Lambda''_z -D''_z & \Lambda''_y-D''_y \\
			-\Lambda''_z+D''_z & J''_y & -\Lambda''_x-D''_x \\
			\Lambda''_y+D''_y & -\Lambda''_x+D''_x & J''_z 
		\end{pmatrix}.
        \label{aNPR:eqn:Ybond}
   		\end{equation} 
	For a $\bm{q} = 0$ magnetic structure, the interaction energy between the Pr(1) and Pr(2) sublattices can be expressed in terms of the average exchange matrix $\mathcal{J}=(\mathcal{J}^{\rm X}+\mathcal{J}^{\rm Y}+\mathcal{J}^{\rm Z})/3$ obtained to have the form   	
	\begin{equation}
		\mathcal{J} = \begin{pmatrix}
			J_x			& 0 			& \Lambda-D \\
			0 			& J_y 	& 0 \\
			\Lambda+D	& 0 			& J_z
		\end{pmatrix}   
		\label{aNPR:calc:effective}
	\end{equation}  
    with $J_\alpha=(J'_\alpha + 2J''_\alpha)/3$, $\alpha\in\left\{x,y,z\right\}$, $\Lambda=(\Lambda_y'+2\Lambda_y'')/3$ and $D=(D_y'+2D''_y)/3$. Note that the only symmetry-allowed off-diagonal terms are $\mathcal{J}_{xz}$ and $\mathcal{J}_{zx}$, which contain DM and symmetric off-diagonal exchange in the $xz$ plane. In this notation $D$ is the average of the $y$-components of the DM vectors on the X, Y and Z bonds, so an effective DM vector along the $b$-axis. 

	A minimal model to stabilise the zero-field magnetic structure proposed in Sec.~\ref{aNPR:sec:mag_sym} has easy-axis exchange anisotropy $J_z>J_x = J_y>0$ and finite DM interaction term $D$, referred to as the XXZ$+D$ Hamiltonian. The $z$-spin components are antiferromagnetically-aligned by the dominant easy-axis exchange $J_z$, and a small canting of the spins towards the $x$ axis by an angle $\phi=\frac{1}{2}\arctan\frac{2D}{J_x+J_z}$ occurs due to the DM interactions. For the Hamiltonian parameters listed in Table~\ref{aNPR:calc:tab:parameters} $\phi\simeq 1.2^\circ$. A finite $\Lambda$ term would rigidly rotate this canted spin arrangement in the $xz$ plane around the $+b$ axis by an angle $\theta=\frac{1}{2}\arctan\frac{2\Lambda}{J_z-J_x}$, however in the following we consider $\Lambda=0$ since the ferromagnetic moment is observed experimentally to be close to the $x$-axis.  

    For concreteness, in the following we perform all mean-field calculations for an XXZ$+D$ model where all bonds have the same exchange matrix ($\mathcal{J}^{\rm X,Y,Z}=\mathcal{J}$), but note that a wider class of models where the individual bond exchange matrices have the general forms in eqs.~(\ref{aNPR:eqn:Zbond})-(\ref{aNPR:eqn:Ybond}) with the same average $\mathcal{J}$ have an identical mean-field behavior as long as their magnetic ground state has $\bm{q}=0$. The equilibrium magnetic structure in a magnetic field $\bm{B}$ applied along a general orientation was determined by finding numerically the optimal orientations of the four spins in the primitive cell that minimize the mean-field energy for the above exchange model augmented by the Zeeman energy $-\sum_i\mu_{\rm B}\mathbf{B}\cdot\mathbb{G}\cdot\mathbf{S}$ where $\mathbb{G}$ is 
    the $g$-tensor, assumed to be the same for the two Pr sites. In order to ensure that the zero-field magnetic structure has the spins of the same sublattice in adjacent layers parallel as in the proposed zero-field magnetic structure in Fig.~\ref{aNPR:fig:magnetic} we also assume a very small ferromagnetic Heisenberg coupling between nearest-neighbor same sublattice sites (Pr1-Pr1 and Pr2-Pr2) in adjacent layers ($J_{c}=-0.001 J_z$). Such a small inter-layer coupling has minimal effects on the phase diagram at finite field. We found that the complete magnetic phase diagram up to saturation could be captured within this primitive cell. For comparison with the experiments we assume $\bm{B}=\mu_0\bm{H}$. The Hamiltonian parameters used are listed in Table~\ref{aNPR:calc:tab:parameters}. The $J_{x,z}$ values are fixed to those proposed in Ref.~\cite{daum2021collective}. The torque is not sensitive to the sign of $D$, only its magnitude, which was chosen such as to reproduce the observed ratio of the torque peak-to-peak bandwidth at 4~T in the $ac^*$ plane in Fig.~\ref{aNPR:fig:torque:fields}(a) and $ab$ plane in Fig.~\ref{aNPR:fig:torque:ab}(c) measured on the same sample. We assume a uniaxial $g$-tensor with unique axis $z$, with values chosen to simultaneously satisfy the following constraints: (i) root-mean-square spherically-averaged $g$-factor (0.995) is close to the previously determined value $0.98$ inferred from powder susceptibility measurements \cite{daum2021collective}, (ii) $g_{x,y} > g_z$ such as to give a simple explanation to the order of the magnetic susceptibilities $\chi_{a,b}>\chi_{c^*}$ at temperatures $\gg T_{\rm N}$ inferred from our angle-dependent torque data in Sec.~\ref{S:torque16T}, and (iii) the model predicts a spin-flop transition field in a range that could be compatible with the high-field torque data to be discussed later in Sec.~\ref{aNPR:sec:high-field}. For concreteness, we choose parameters such that the (numerically) calculated zero-temperature spin-flop field comes out to be $\mu_0H_{\rm SF}(T=0)\simeq24.3$~T for $H\parallel c^*$, near the zero-temperature value obtained from one plausible extrapolation of the experimental phase boundary plotted in Fig.~\ref{aNPR:fig:nijmegen:c:temperature}(d) (dash-dotted line), noting that other extrapolations could also be compatible with the existing high-field data. For simplicity, we also neglect the very small $g$-factor anisotropy in the $ab$ plane inferred from the paramagnetic torque data and set $g_x=g_y$. 
    
    The above Hamiltonian parameters are under-constrained and should be regarded as representative for the purpose of illustrating the level of agreement that can be obtained with the torque data. We note that the zero-field canted spin structure is independent of the $g$-tensor and all key features of the low-field torque angular dependence in the ordered phase (stable and unstable equilibrium positions, presence or absence of torque jumps when rotating field, overall torque angular shape and low-field scaling) are essentially determined by the exchange part of the spin Hamiltonian and are relatively robust against reasonable variations of the $g$-tensor. However, the magnetic moments and the torque magnitudes do depend quantitatively on the $g$-tensor, which in our parameterization depend on the value assumed for the extrapolated  $\mu_0H_{\rm SF}(T=0)$. 
    
    We note that a recent study of the crystal-field transitions in \aNPR \cite{Onuorah2024} proposed a $g$-tensor anisotropy of the opposite sign to what we considered, $g_{x,y} < g_z$. In this scenario, assuming that the high-temperature susceptibilities at $T\gg T_{\rm N}$ are dominated by the $g$-factor anisotropies, one would expect upon increasing temperature a change in the sign of the susceptibility anisotropy from $\chi_{a,b}>\chi_{c^*}$ in the ordered phase to $\chi_{a,b}<\chi_{c^*}$ at $T\gg T_{\rm N}$. However, our angle-dependent torque data in both the $ac^*$ plane [see Fig.~\ref{aNPR:fig:torque:temperature}(a)] and the $bc^*$ plane [see Fig.~\ref{aNPR:fig:torque:bc}(a)] show no sign change in the torque angular dependence over the temperature range 2-100~K. In particular, in the high-temperature paramagnetic regime the stable and unstable equilibrium positions imply $\chi_{a,b}>\chi_{c^*}$, the same as we deduced for the ordered phase. Future studies focused on determining the $g$-tensor directly would be very beneficial for making progress towards refining the Hamiltonian parameters and finding a consistent explanation of all experimental observations in terms of a common microscopic model.  
    
    \begin{figure*}[tb]
		\includegraphics[width=0.8\textwidth]{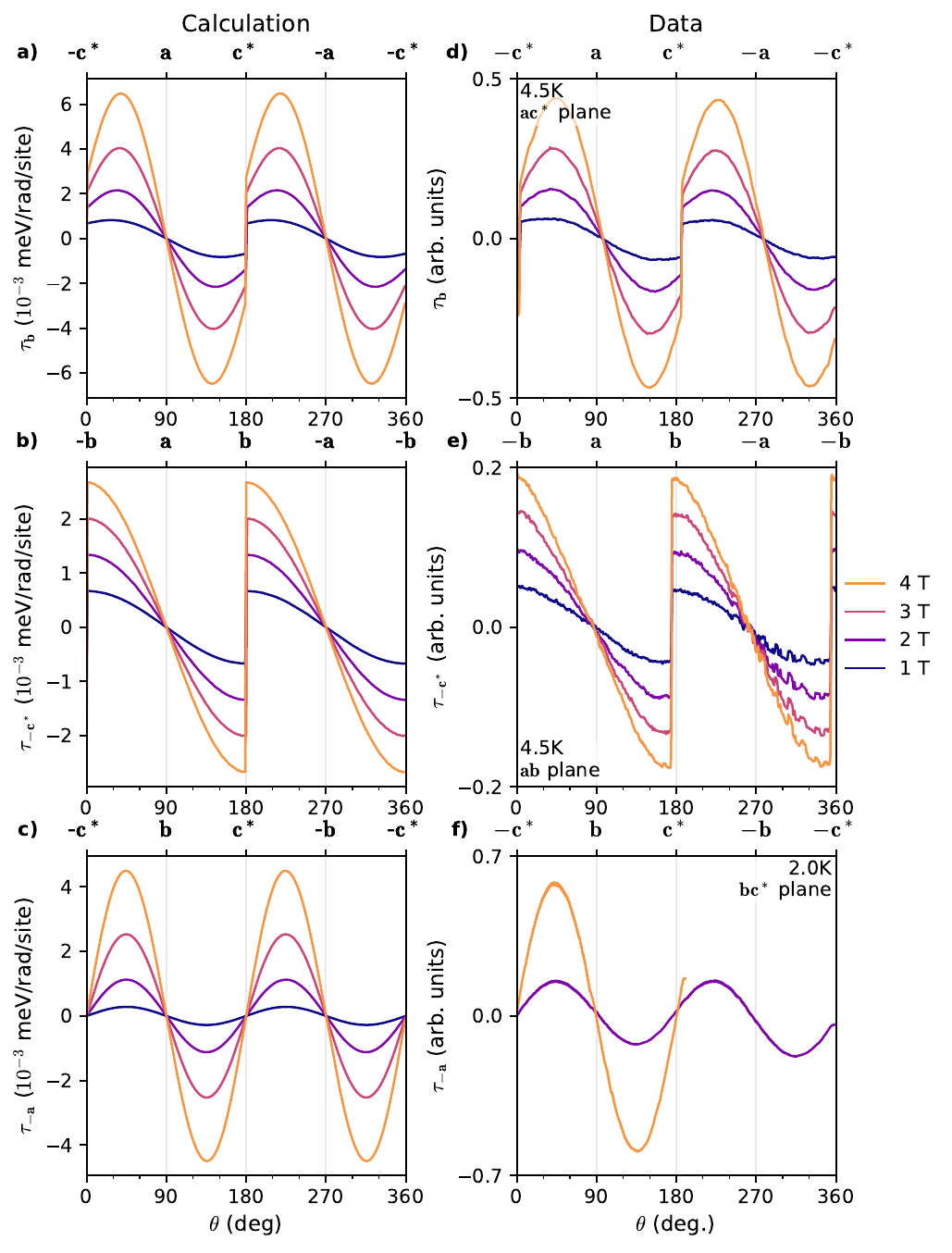}\centering
		\caption[Angular dependence of the magnetic torque in orthogonal crystal planes.]{
        Angular dependence of the magnetic torque, (left) calculations vs. (right) data measured at 4.5~K, for field in the three orthogonal crystallographic planes. Curves are color-coded by the field magnitude. Top axes labels in each panel indicate when the applied field is oriented along reference crystallographic axes. The calculated torque (left) using parameters in Table~\ref{aNPR:calc:tab:parameters} is in absolute units, the measured torque (right) is in arbitrary units, with the same scale between (d) and (e), but different from (f), as a different sample was measured in (d,e) compared to (f).
        }
		\label{aNPR:fig:calc:torque}
	\end{figure*}
    
	\begin{table}[tb]
		\centering
		\caption{\label{aNPR:calc:tab:parameters} Hamiltonian parameters defined in eq.~(\ref{aNPR:calc:effective}) and used in the mean-field calculations. 
        }
		\begin{tabular}{ccc}
			\hline 
            \hline
            $g_x,g_y$ & \quad\quad\quad &1.09\\ 
            $g_z$ & & 0.77 \\
			$J_x,J_y$ && 1.06~meV\\ 
            $J_{z}/J_x$ && 1.22 \\
			$D$ && 0.05~meV \\
			$\Lambda$ &&  0\\
			\hline
            \hline
		\end{tabular}
	\end{table} 
    
    \subsection{Torque angular dependence}
    
	Fig.~\ref{aNPR:fig:calc:torque} shows a direct comparison between the calculated (left) and observed (right) torque angular dependence in the three crystallographic planes using the Hamiltonian parameters in Table~\ref{aNPR:calc:tab:parameters}. Note that all key features seen experimentally are reproduced, including the torque sign, positions of stable/unstable equilibrium, presence of prominent torque jumps for field in the $ac^*$ and $ab$ planes, and absence of such jumps in the $bc^*$ plane. In each of the three planes, the scaling of the torque signal with the field magnitude (color-coded curves) is also reproduced. 

	\begin{figure}
		\includegraphics[width=0.45\textwidth]{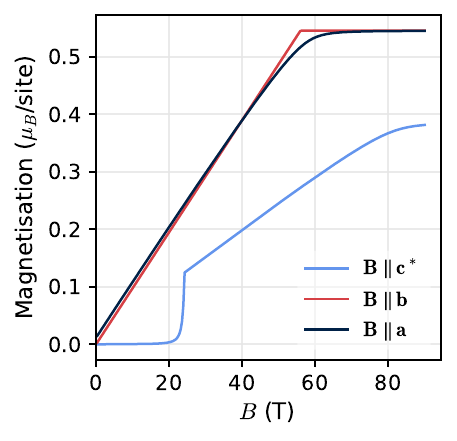}
        \centering
		\caption[Calculated magnetization curves.]{Calculated magnetization curves for field along different crystallographic directions (color-coded) for the minimal Hamiltonian model with parameters in Table~\ref{aNPR:calc:tab:parameters}.}
		\label{aNPR:fig:calc:sweeps}
	\end{figure}

    \subsection{Magnetization curves}
	The calculated magnetization curves along the three reference crystallographic directions are plotted in Fig.~\ref{aNPR:fig:calc:sweeps}. Field along the two-fold $b$-axis is normal to the plane of the ordered spins in zero field (see Fig.~\ref{aNPR:fig:magnetic}), therefore upon increasing field spins cant progressively towards the field direction resulting in a magnetization curve (red) that starts at zero and is linear in field up a sharp transition to saturation. Below saturation there are two magnetic domains related by the 2-fold rotation along $b$ that passes through each Pr site.   
    
    The behavior for field along the other two orthogonal directions is more complex. In particular, an infinitesimal field along the $a$-axis selects one of the two zero-field degenerate magnetic domains (the one with ferromagnetic moment along the applied field direction), so the magnetization curve (black) starts from the finite value of the spontaneous zero-field magnetization, then increases linearly upon increasing field as the spins that were initially canted only a small angle away from the $\pm c^*$ axes towards $a$, now cant progressively more upon increasing field. Note however that the curve smoothly turns over upon reaching saturation with no sharp transition encountered, only a crossover. This is because at any finite field along $+a$ the DM interaction selects a unique ground state, the one shown in Fig.~\ref{aNPR:fig:magnetic} but with progressively more canting towards $a$ upon increasing field, with the Pr(1) spins down-left and Pr(2) spins up-left (the magnetic structure with the signs of the antiferromagnetically-coupled components swapped between the Pr(1) and (2) sites has a higher energy due to the DM interaction). 
    
    Finally, field along $c^*$ shows most complex behavior as this is along the direction of the antiferromagnetically-aligned components in the zero-field magnetic structure. Upon increasing field from zero the canted arrangement of the Pr(1) and (2) spins gradually rotates with both spins confined to the $ac^*$ plane. The rotation occurs in opposite senses for the two magnetic domains, around $\mp b$ for the domain with spontaneous ferromagnetic polarization along $\pm a$, such that the net ferromagnetic moment $\mathbf{m}$ rotates monotonically towards the field, with the rotation rate increasing as the field increases. For a given magnetic domain, as spins (1) and (2) rotate the relative angle between them decreases (the spin that was initially aligned opposite to the field rotates faster) such that $\mathbf{m}$ increases in magnitude and rotates at the same time, until at the critical spin-flop field $B_{\rm SF}$ the net ferromagnetic moment is exactly along the field direction and both magnetic domains end up in the same state, with Pr(1) and (2) spins canted away symmetrically from $c^*$ towards $+a$ and $-a$, respectively. Note that: (i) the ground state in the spin-flop state is unique, i.e. the state with the antiferromagnetically-aligned components swapped between the Pr(1) and (2) sites is higher in energy due to DM interaction, and (ii) the transition at $B_{\rm SF}$ is found numerically to be continuous, with the DM interactions allowing the two canted AFM states to evolve continuously into the unique spin-flop ground state. The order parameter for this transition is the spontaneous transverse moment $m_x$ tuned continuously to 0 as $B$ approaches $B_{\rm SF}$ from below. Upon further increasing field above $B_{\rm SF}$ the magnetization increases linearly as the spins cant progressively more towards the $c^*$ axis. The magnetization curve smoothly bends over upon reaching saturation in a crossover rather than a sharp phase transition, again due to the DM interactions having already selected a unique (single domain) spin-flop ground state.       

	\begin{figure}
		\includegraphics[width=0.5\textwidth]{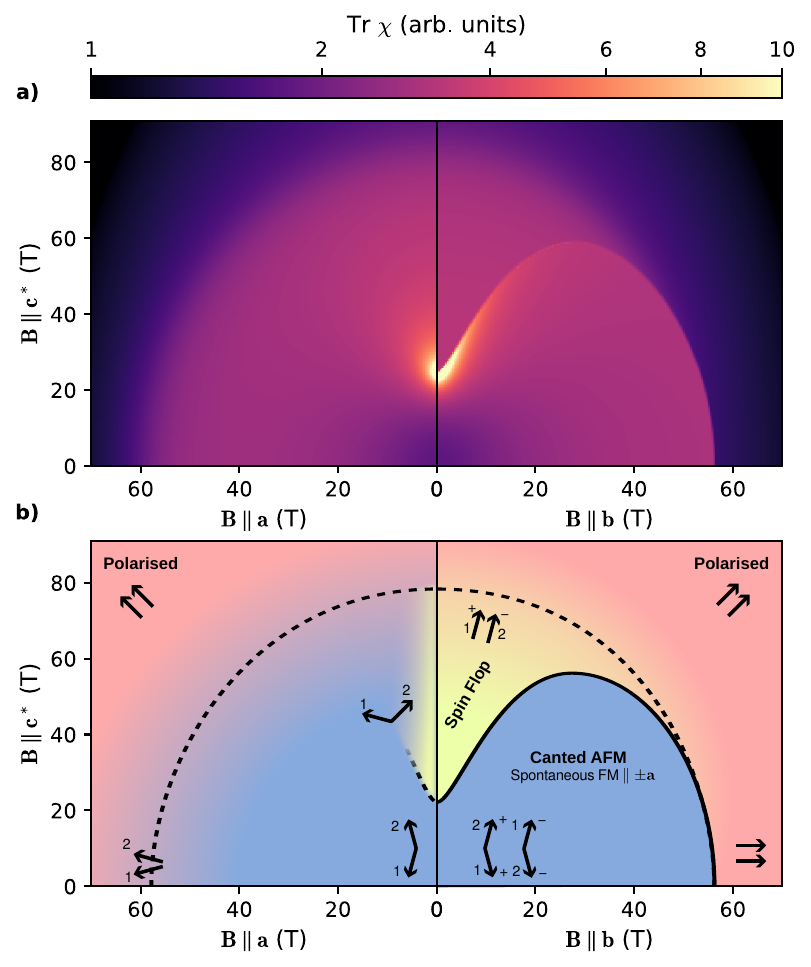}\centering
		\caption[Calculated phase diagram of \aNPR]{
        (a) Calculated field dependence of the trace of the differential susceptibility tensor Tr$\chi$ represented as a color map for field in the $ac^*$ (left) and $bc^*$ plane (right). (b) Inferred magnetic phase diagram, where shaded regions indicate reference magnetic ground states described in the text, dashed lines over a smoothly varying color shading are smooth crossovers and solid lines separating distinct colors correspond to sharp phase transitions. The Hamiltonian parameters are given in Table~\ref{aNPR:calc:tab:parameters}. The magnetic structure in various parts of the phase diagram is indicated by diagrams of two black arrows, which show the projection of the Pr(1) and (2) spins onto the field varying plane at that point, with $\pm$ signs indicating the spins' projection onto the out-of-plane axis, left side $b$, right side $a$.}
		\label{aNPR:fig:calc:phase}
	\end{figure}

    \subsection{Magnetic phase diagrams}
	Fig.~\ref{aNPR:fig:calc:phase}(b) shows the calculated magnetic phase diagram for the minimal model for field in the $ac^*$ (left) and $bc^*$ (right) planes, where solid (dashed) lines indicate sharp phase transitions (smooth crossovers). The location of phase boundaries and crossovers was obtained from an analysis of the trace of the differential susceptibility tensor in the plane where the magnetic field is varied, i.e. ${\rm Tr} \chi = \chi_{\mu\mu}+\chi_{\nu\nu}$, where 
	\begin{equation}
		\chi_{\eta\eta} = -\frac{\partial^2 \mathcal{F}}{\partial B^2_\eta}, \quad \eta=\mu, \nu\label{aNPR:eqn:diffsus}
	\end{equation}
	with $\mathcal{F}$ the free energy, which equals the ground state energy at $T=0$. The subscripts $\mu\nu$ are $xz$ ($yz$), respectively for Fig.~\ref{aNPR:fig:calc:phase}(a) left (right) panel, which shows a color plot of ${\rm Tr} \chi$ over the same field range as the magnetic phase diagrams in panel (b). Sharp phase transitions are expected to be manifested in clear non-analytic anomalies, whereas crossovers correspond to rapid, but smooth variations.  
    
    For field in the $ac^*$ plane (left panel) there is no sharp phase transition upon increasing field except the transition to the spin-flop state at $B_{\rm SF}$ for field precisely along $c^*$ (vertical axis direction). For field below $B_{\rm SF}$ any finite field component along $a$ selects a unique ground state out of the two zero-field degenerate canted AFM states and upon increasing field the canted Pr(1) and (2) arrangement evolves smoothly such as to increase the magnetization along the field. Consequently the approach to the field-polarized state for any field orientation in the $ac^*$ plane is then a smooth crossover (dashed line).  
    
	For field in the $bc^*$ plane [right panel of Fig.~\ref{aNPR:fig:calc:phase}(b)] there is a sharp phase boundary (solid line) of continuous transitions bordering the canted AFM phase (blue shaded region) and the spin-flop region (yellow shading) at intermediate fields. This is manifested in a sharp drop in ${\rm Tr} \chi$ upon increasing field. We have also verified numerically that the {\em transverse} susceptibility $\chi_{xx}$ diverges upon crossing this phase boundary coming from higher fields, as expected since the spin-flop state has no ferromagnetic polarization along the $a$-axis and the canted AFM phase has two degenerate magnetic domains with spontaneous ferromagnetic polarization along the $\pm a$ axis. Furthermore, we have also verified numerically (using the \textsc{SpinW} package \cite{spinw}) that the linear spin-wave spectrum is in general gapped in all regions of the phase diagram but becomes gapless at the zone center ($\mathbf{q}=0$) for fields precisely on the solid line phase boundary between the canted AFM and the spin-flop region, as expected for a continuous transition between those two phases when ferromagnetic fluctuations become critical. There is only a crossover (smooth variation in ${\rm Tr} \chi$) between the spin-flop region and the field-polarized state. Tilting the field out of the $bc^*$ plane to have a finite component along $a$ lifts the degeneracy of the two canted AFM states, makes the ground state unique and transforms the nominal transition line between the canted AFM and spin-flop region into a crossover.  
           
    Note that the phase diagram has significant qualitative differences compared to that of the pure easy-axis XXZ Hamiltonian with no DM interactions. In this case the upper dashed line becomes a sharp phase transition to magnetization saturation, for fields below this line there are always two degenerate magnetic domains obtained by reversing the antiferromagnetically-aligned spin components, the phase diagrams in the $ac^*$ and $bc^*$ planes are the same by symmetry, and the spin-flop transition is first order and confined to fields strictly along $c^*$, with any infinitesimally small transverse field component converting this transition into a smooth crossover \cite{spin-flop}. In contrast, the addition of DM interactions with $\mathbf{D}\parallel \mathbf{b}$ removes completely one magnetic domain for field in the $ac^*$ plane away from $c^*$ and makes the approach to magnetization saturation a smooth crossover. For field in the $bc^*$ plane, DM interactions stabilize a sharp, continuous phase transition boundary between the low-field canted AFM phase with spontaneous ferromagnetic polarization along $\pm a$, and the intermediate-field spin-flop region, with a smooth crossover to magnetization saturation upon further increasing field for a general field orientation away from the $b$-axis. By comparing the calculated phase diagrams for variable $D$, we have found that the gapless line in the $bc^*$ plane evolves continuously as $D \to 0$ into a gapless field path for the pure XXZ model. This path consists of the gapless spin-flop phase for field along $c^*$ extending from the spin-flop transition to saturation along $c^*$, followed by the saturation transition phase boundary as the field is rotated towards $b$.
    
    \subsection{Alternative $J\Gamma+D$ Hamiltonian}
	For completeness, we note that an alternative to the XXZ$+D$ model discussed up to now, a so-called $J\Gamma+D$ model with suitably chosen parameters would also have the same ground state magnetic structure {\em and} the same behavior in applied field. 
    The $J\Gamma$ model is a special case ($K=0$) of generalized Kitaev $JK\Gamma$ models \cite{rau2014generic} explored in the context of honeycomb iridates and $\alpha-$RuCl$_3$, defined as 
    \begin{equation}
		\mathcal{H}_{JK\Gamma} = \sum_{\left<ij\right> \in \alpha\beta(\gamma)} J \bm{S}_i\cdot\bm{S}_j + K S_i^\gamma S_j^\gamma + \Gamma(S_i^\alpha S_j^\beta + S_i^\beta S_j^\alpha),\nonumber
	\end{equation}
	where the sum is over the nearest-neighbor bonds. Each bond has an associated Kitaev axis $\gamma \in \left\{\sf{x,y,z}\right\}$ and $\alpha\beta(\gamma)$ refer to the orthogonal plane to this axis. The Kitaev axes  $\sf{xyz}$ (denoted by SansSerif symbols) are related to the crystallographic-based $xyz$ axes via 
    \begin{equation}
        \left(\begin{array}{c}        
        \bm{\mathsf{\hat{x}}} \\
        \bm{\mathsf{\hat{y}}} \\
        \bm{\mathsf{\hat{z}}}
        \end{array}\right)
            =P 
        \left(\begin{array}{c}
            \mathbf{\hat{x}} \\
            \mathbf{\hat{y}} \\
            \mathbf{\hat{z}}
            \end{array}\right), \quad
    P = 
		\begin{pmatrix}
			-1/\sqrt{6} & 1/\sqrt{2} & 1/\sqrt{3} \\
			-1/\sqrt{6} & -1/\sqrt{2} & 1/\sqrt{3} \\
			2/\sqrt{6} & 0 & 1/\sqrt{3}  
		\end{pmatrix},
        \nonumber
    \end{equation}
    where $\mathbf{\hat{x}}$ indicates the unit vector along the $x$-axis and so on. The above definition is chosen such that in an idealized structure where the Pr honeycombs have three-fold rotational symmetry around $c^*$ and the PrO$_6$ octahedra are cubic, the $\gamma$ direction of each bond is then normal to the local Pr(1)-O$_2$-Pr(2) superexchange plane of that bond.  
    The effective interaction between the Pr(1) and (2) sublattices in a $\mathbf{q}=0$ structure, when expressed in the $xyz$ coordinate frame, is obtained to be of the same form as eq.~(\ref{aNPR:calc:effective}) with 
    \begin{equation}
    J_x=J_y=J - \frac{\Gamma}{3}, \quad {\rm and} \quad J_z=J + \frac{2\Gamma}{3}.
    \label{aNPR:eqn:JG}
    \end{equation}
    
    We have found numerically that for Hamiltonian parameters within some range of those listed in Table~\ref{aNPR:calc:tab:parameters} (weakly Ising-like XXZ and small $\Gamma$ at finite $D$) the minimum energy state of both the XXZ$+D$ model and the $J\Gamma+D$ model (using the parameter mapping in eq.~(\ref{aNPR:eqn:JG})) is a $\mathbf{q}=0$ structure in {\em any} applied field. This is always the case for the XXZ$+D$ model with the same exchange matrix on every bond. However, in the $J\Gamma+D$ model the bond-dependent $\Gamma$ interactions prefer different spin orientations on the different bonds ($\bm{\mathsf{\hat{x}}}+\bm{\mathsf{\hat{y}}}$ for the Z bond, permuted cyclically for the other bonds), the magnetic structure is selected as a compromise between the different bonds and this opens possibilities for more complex structures with $\bm{q}\neq 0$ in the phase diagram for large $\Gamma$ \cite{Suzuki2021}; that regime is relatively far from parameters corresponding to Table~\ref{aNPR:calc:tab:parameters}, for which the magnetic ground state has $\bm{q}=0$ in any field. In this case, even though microscopically distinct, the two models are indistinguishable by their static magnetometry behavior within the present $T=0$ mean-field description. In particular, their calculated torque angular dependence in Fig.~\ref{aNPR:fig:calc:torque}(left column), magnetization curves in Fig.~\ref{aNPR:fig:calc:sweeps} and magnetic phase diagrams in Fig.~\ref{aNPR:fig:calc:phase} are identical. The equivalence only relates to the ground state energy and implicitly the static properties; the two models could in principle be distinguished by their spin-wave spectra. 

    \begin{figure*}[!htb]
	\includegraphics[width=0.99\textwidth]{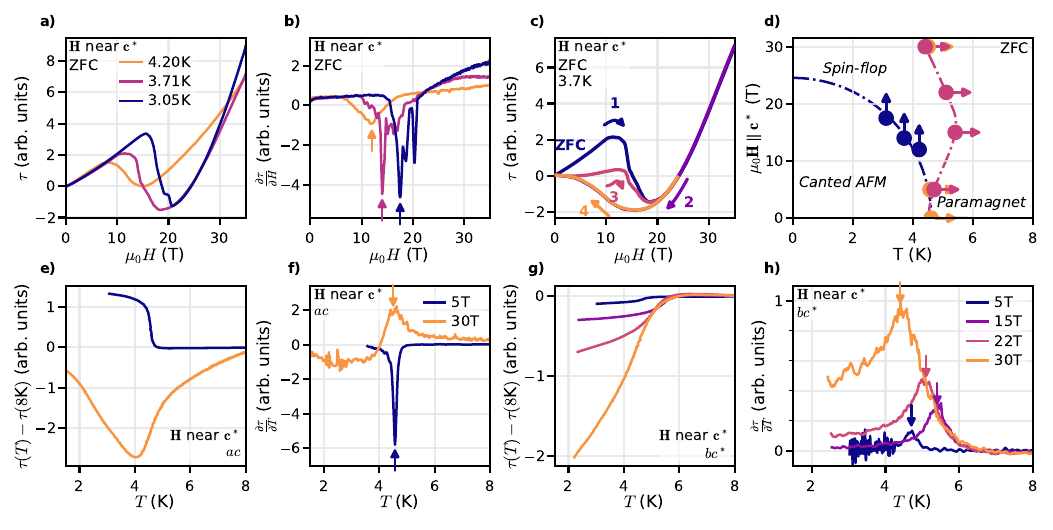}
        \centering
		\caption[Field dependence of magnetic torque up to 35T.]{(a) Torque as a function of field magnitude for field applied close (within a few degrees) to the $c^*$ axis in the $ac^*$ plane at various temperatures (color-coded). Curves are the raw cantilever voltage, assumed to be proportional (up to an undetermined sign) to the torque component along the $b$-axis. (b) First derivative of the curves in (a). Overlaid vertical arrows indicate location of prominent peaks attributed to phase transitions/crossovers. (c) Hysteresis of the torque for zero-field-cooled (ZFC) upsweep (step 1, blue), and subsequent downsweep (step 2), upsweep (step 3), downsweep (step 4). Arrows show sweep direction for each curve, numbers show order in the measurement sequence. (d) Temperature-field phase diagram for field close to the $c^*$ axis, showing the spin-flop, paramagnetic and canted antiferromagnetic phases. Filled symbols indicate anomaly points extracted from scans along the direction of the arrows, ZFC field upsweeps (blue symbols, up arrows), and temperature scans (yellow/purple symbols and horizontal arrows) extracted from the data in (e,f) and (g,h), respectively. Dash-dotted lines are guides to the eye to indicate phase transitions/crossovers. The error bars on the data points are smaller than the size of the symbols. Temperature dependence of (e, g) magnetic torque $\tau$ and (f, h) its first derivative $\frac{\partial \tau}{\partial T}$ for various fixed fields (color-coded) applied close (within a few degrees) to the $c^*$ axis in the $ac^*$ (e, f) and $bc^*$ plane (g, h). Torque curves in (e,g) are offset by their value at 8~K and the torque absolute sign is undetermined.}
        \label{aNPR:fig:nijmegen:c:temperature}
	\end{figure*}
    \section{DC Field Torque Magnetometry up to 35~T}
    \label{aNPR:sec:high-field} 
	
    Additional torque measurements were performed in DC fields up to 35~T at the HFML in Nijmegen to search for evidence of field-induced transitions. 
    Fig.~\ref{aNPR:fig:nijmegen:c:temperature} shows field upsweeps, for field oriented close to the $c^*$ axis in the $ac^*$ plane, between 3 and 4.2~K. Below 9~T, the torque curves at those temperatures overlap and the torque is approximately linear in field. This linear relationship matches the behavior seen at low field previously (in Fig.~\ref{aNPR:fig:torque:fields} for field close to $c^*$, $\tau_b\approx\Delta_1\sim H$). At higher fields, clear non-monotonic anomalies are apparent and become steeper and move to higher field upon cooling. We attribute those features to proximity to the field-induced transition from the low-field canted AFM phase to the higher-field spin-flop state predicted by the minimal model for field strictly along $c^*$ or tilted slightly towards the $b$-axis. We define the experimental anomaly field as the location of the steepest gradient. We focus on the systematics of the most prominent peak in the first derivative $\frac{\partial\tau}{\partial H}$ shown in panel (b) (arrows) and we neglect other small features such as additional shoulders or smaller peaks. The locations of those prominent anomalies in a field-temperature phase diagram are plotted in panel (d) (blue dots). 
    
    The above field-induced anomalies display strong hysteresis effects at low temperatures. This is illustrated in panel (c), which compares the field dependence of torque for two consecutive field upsweeps (steps 1 and 3) and downsweeps (2 and 4). Above $\sim 20$~T all four curves overlap, but significant hysteresis is observed at lower field, between the two downsweeps (steps 2 and 4, which do overlap) and the two upsweeps (1 and 3), which also differ between them, with all four curves overlapping again at zero field. We attribute the observed hysteresis to metastability effects associated with proximity to the field-induced spin-flop phase transition, although a detailed understanding is outside the scope of the present study as it requires going beyond the minimal model in Table~\ref{aNPR:calc:tab:parameters}, which predicts a sharp, continuous spin-flop transition for field strictly along $c^*$ or tilted towards the $b$-axis, but which converts to a smooth crossover into the spin-flop regime if the field has a finite $a$-axis component.  
	
	Temperature sweeps in fixed fields were performed to explore the regions of the phase diagram where the phase boundary is close to field-independent, i.e. approximately vertical in Fig.~\ref{aNPR:fig:nijmegen:c:temperature}(d). Technical aspects of the thermometer calibration for these measurements are presented in Appendix~\ref{aNPR:sec:calib}. Fig.~\ref{aNPR:fig:nijmegen:c:temperature}(e-f) show the observed temperature dependence of the torque at fixed field of 5 and 30~T, raw data (e) and first derivative (f). At 5~T, there is a sharp upturn in the raw torque (sharp peak in the derivative) at 4.56(1)~K, associated with the onset of magnetic order and shown with a yellow point in the phase diagram in Fig.~\ref{aNPR:fig:nijmegen:c:temperature}(d). At 30~T, a clear anomaly is also observed, manifested in a broad peak in the torque derivative near 4.5(1)~K. We interpret this broad peak as the expected crossover from the low-temperature spin-flop to the high-temperature paramagnet. Panels (g-h) show analogous measurements for field also close to the $c^*$ axis, but in the $bc^*$ plane. The torque derivative shows a broad peak at all measured fields, and the location of the peak in temperature is strongly field dependent and non-monotonic, increasing up to 5.4(1)~K at 15~T then decreasing again upon increasing field. The anomalies extracted from temperature scans are plotted as purple symbols in the phase diagram in Fig.~\ref{aNPR:fig:nijmegen:c:temperature}(d).

    As already discussed in the previous Sec.~\ref{aNPR:sec:mean-field}, for fields above the spin-flop transition the magnetic ground state is expected to be unique (single domain) due to the presence of DM interactions, which make the alternative magnetic structure with swapped signs of the antiferromagnetically-aligned components higher in energy. Because of this, the spin-flop state has the same symmetry as the field-polarized state, so no symmetry-breaking phase transition is expected between them upon increasing field, consistent with the upper dashed line in Fig.~\ref{aNPR:fig:calc:phase}(b). For the same reason, no sharp symmetry-breaking transition is expected upon increasing temperature from the spin-flop region into the high-temperature paramagnet. This might explain why the torque anomalies associated with temperature scans at high field are rather broad, as expected from a smooth crossover rather than a sharp anomaly associated with a symmetry-breaking phase transition. 

    \section{Discussion}
    \label{aNPR:sec:discussion} 
    
    The zero-field magnetic order consisting of an antiferromagnetic N\'{e}el pattern with a weak spontaneous ferromagnetism is quite different from the magnetic order observed in other honeycomb Kitaev candidate materials with edge-sharing octahedra such as $\alpha$-RuCl$_3$ \cite{Cao2016} and \NaIrO \cite{Liu2011}(antiferromagnetic zigzag order) and $\alpha$-\LiIrO \cite{Williams2016}(incommensurate spiral order). In particular the weak ferromagnetism is a distinct feature of \aNPR that we attribute to the fact that its crystal structure  has two inequivalent Pr sites arranged alternatingly on the sites of the honeycomb lattice. This breaks the inversion symmetry of $b$-axis bonds and allows for DM interactions, which we propose are responsible for the canting of the magnetic moments and the observed weak spontaneous ferromagnetism along the in-plane $a$-axis. In the other layered honeycomb materials listed above all the honeycomb sites correspond to a single crystallographic site, the $b$-axis bonds are inversion-symmetric, which rules out DM interaction on those bonds, so removes the mechanism for moments canting and inducing weak spontaneous ferromagnetism. 
    
    The presence of DM interactions is expected to alter qualitatively the magnetic phase diagrams in applied field. A non-trivial prediction of the minimal model is that a sharp phase boundary (black solid line) exists for field in the $bc^*$ plane, normal to the $a$ direction of the spontaneous zero-field magnetization. This transition line connects the spin-flop transition for field along $c^*$ to the transition to magnetic saturation for field along $b$, see Fig.~\ref{aNPR:fig:calc:phase}~(b)~(right). Such a phase boundary line does not exist in the absence of DM interactions, and is also not predicted for fields in the $ac^*$ plane that contain the direction of the spontaneous zero-field magnetization, see Fig.~\ref{aNPR:fig:calc:phase}~(b)~(left), which could be tested by magnetometry and/or heat capacity experiments at high fields. In terms of symmetry considerations, the two-fold magnetic ground state degeneracy in zero field is guaranteed by time-reversal symmetry, and also by the $2_y$ two-fold symmetry axes of the $C2/c$ structure that pass through each Pr site. In field applied along the $b$-axis, even though time-reversal symmetry is broken, the $2_y$ rotations survive and protect the two-fold ground state degeneracy. The two-fold ground state degeneracy for field away from the $b$-axis into the $bc^*$ plane (normal to the zero-field magnetization) is due to an accidental symmetry of the minimal XXZ$+D$ model we have considered, not due to a symmetry operation of the $C2/c$ crystal structure. The two degenerate ground states are related by swapping the spins at the two Pr sites and reversing their $S_x$ components, equivalent to applying {\em for each layer separately} a $2'_x$ rotation through the middle of a $b$-axis bond (2-fold rotation along the $a$-axis of the spontaneous zero-field magnetization, followed by time reversal). This operation leaves the minimal model Hamiltonian invariant for field in the $bc^*$ plane. Other subleading symmetry-allowed Hamiltonian terms such as slightly different $g$-tensors at the two inequivalent Pr sites, would lift the degeneracy of the two states connected by the above operation, make the ground state unique for any field tilted away from the $b$-axis into the $bc^*$ plane for the minimal model and transform the nominal transition line between the canted AFM and the spin-flop region into a crossover. However, if those subleading corrections are sufficiently weak, then non-monotonic features may still be observed in the thermodynamic quantities upon crossing the putative transition line expected for the minimal model.

    We have also verified numerically that adding other small perturbations to the minimal model, such as a small symmetric off-diagonal exchange $\Lambda$ in eq.~(\ref{aNPR:calc:effective}), making $g_x\neq g_y$ and/or adding a small off-diagonal $g_{xz}$ term, all symmetry-allowed by the $C2/c$ crystal structure, preserve the line of continuous transitions, but finite $\Lambda$ and/or $g_{xz}$ can in general move it from the $bc^*$ plane to a nearby $bw$ plane obtained by rotation by a small angle around $b$. We find the same symmetry-breaking physics in this rotated $bw$ plane: below the line of continuous transitions there is spontaneous ferromagnetism along the direction normal to this plane $\bm{u}=\bm{b}\times\bm{w}$ and all the qualitative features described earlier are preserved.
    
    Strong Kitaev and other bond-dependent anisotropic interactions have been proposed to explain the magnetic order patterns in the above honeycomb materials with Ru$^{3+} 4d^5$ and Ir$^{4+} 5d^5$ moments, whereas the dominant interactions in \aNPR between \pr $4f^1$ moments appear to be quite different, described instead by an easy-axis exchange model. Microscopically, the easy-axis anisotropic exchange could arise either from a conventional bond-independent XXZ exchange with the unique axis along $c^*$, or equivalently from a bond-dependent off-diagonal symmetric exchange $\Gamma$ that couples the spin components in the local Pr-O$_2$-Pr superexchange plane of the bond, with {\em identical} static behavior of the XXZ$+D$ and $J\Gamma+D$ models at the mean-field level for the parameter regime relevant to \aNPR. We note that these two types of models have been distinguished in favour of a $J\Gamma$-type model in the related hyperhoneycomb \bNPR polymorph \cite{okuma2023noncollinear}. The crystal structure is different from that of \aNPR, also made of edge-sharing PrO$_6$ octahedra, but where additional bond rotations lead to a three-dimensionally coupled hyperhoneycomb lattice for the Pr ions. In this lattice the local honeycomb plane that defines the local three-fold coordination at each Pr site is not the same for all sites as in the case of the planar honeycomb, but alternates in orientation between sites connected by vertical $c$-axis bonds; this feature allows an easier distinction between XXZ and $J\Gamma$ type models on that lattice already at the mean-field level.         
    
    \section{Conclusions}
    \label{aNPR:sec:conclusion} 
    In summary, we have performed torque magnetometry on untwinned single crystals of \aNPR up to 35~T. From the observed torque angular dependence for field in three orthogonal crystallographic planes ($ac^*$, $bc^*$ and $ab$) we have proposed a minimal model for the low-field magnetic structure with dominant antiferromagnetism along the $c^*$ axis and a small spontaneous ferromagnetic moment along the $\pm a$ axis. We have explained how this magnetic structure is decomposed in terms of symmetry-allowed magnetic basis vectors that belong to a single irreducible representation. We have proposed a minimal anisotropic exchange model with easy-axis XXZ antiferromagnetic exchange and symmetry-allowed Dzyaloshinskii–Moriya (DM) interactions with an effective DM vector along the two-fold $b$-axis and showed that at the mean-field level the full torque angular dependence in all three crystallographic planes measured could be well reproduced, including the field scaling up to 4~T. We have shown that the mean-field calculated magnetic phase diagrams for the minimal model for field in the $ac^*$ and $bc^*$ planes are qualitatively different due to the presence of DM interactions, which convert some phase transitions to magnetization saturation into smooth crossovers and also stabilize a phase boundary line connecting the spin-flop transition in field along $c^*$ to the transition to magnetization saturation in field along $b$. These two transitions are not connected in the absence of DM interactions. In torque measurements as a function of sweeping field up to 35~T near normal to the layers we have observed clear hysteretic anomalies near $\mu_0H_{\rm SF}\simeq17$ T at 3~K, which we attribute to proximity to the field-induced spin-flop transition expected for field along $c^*$. Temperature sweeps at fixed field up to 30~T reveal an anomaly as a function of temperature, manifested as a broad peak in the first derivative, which we attribute to the expected crossover between the spin-flop and paramagnetic regions of the phase diagram.     
    
    \begin{acknowledgments} 
    This research was partially supported by the European Research Council under the European Union’s Horizon 2020 research and innovation programme Grant Agreement Number 788814 (EQFT). AIC acknowledges support from EPSRC grants EP/I004475/1	and EP/I017836/1. For the high-field measurements in Nijmegen we acknowledge support from HFML-FELIX, member of the European Magnetic Field Laboratory (EMFL), and from the Engineering and Physical Sciences Research Council (EPSRC, UK) via its membership of the EMFL (grant EP/N01085X/1). Crystal structure figures were made using \textsc{vesta} \cite{Momma2011}.
    \end{acknowledgments}

    \begin{figure*}[tb]
	\includegraphics{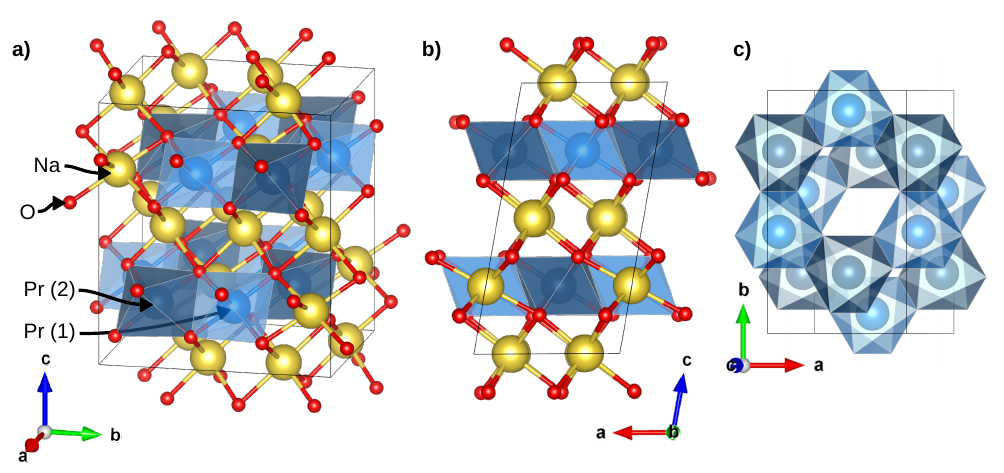}\centering
	\caption[Model showing the crystal structure of \aNPR.]{Monoclinic $C2/c$ crystal structure of \aNPR showing the honeycombs of edge-sharing PrO$_6$ octahedra (blue shading). The two honeycombs per unit cell at $z=1/4$ and $3/4$ are related by a $c$-glide (mirror in the $ac$ plane followed by translation by $\bm{c}/2$). Light and dark blue spheres indicate the two structurally inequivalent \pr ions, yellow and red spheres are \na and \ox ions, respectively. The thin lines show the outline of the monoclinic unit cell. (a) Perspective view. (b) Projection on the $ac$ and (c) $ab$ plane, where in the latter the Na and O ions are omitted for clarity.}
	\label{aNPR:fig:structure}
    \end{figure*}
    
\appendix
\section{X-ray structural studies}\label{aNPR:sec:xrays}    
 
    \begin{figure}[tb]
	\includegraphics{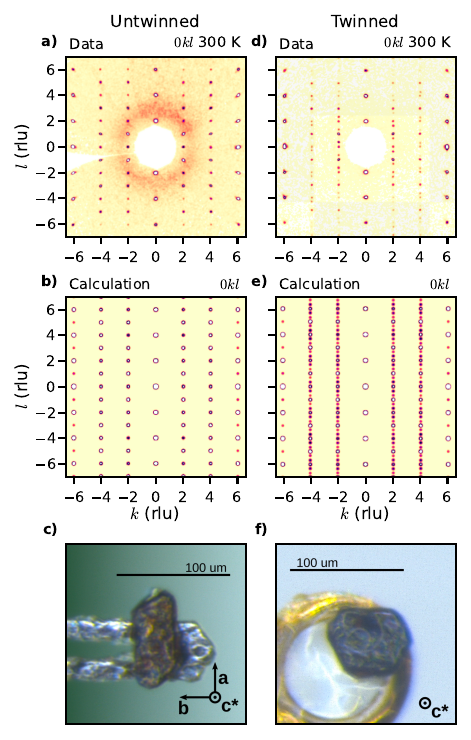}\centering
	\caption[X-ray diffraction pattern and crystal morphology of \aNPR.]
    { Observed (top) and calculated (middle) room-temperature X-ray diffraction patterns in the $(0kl)$ plane for an untwinned crystal (left) and a twinned sample (shown in (f)) with two additional grains rotated by $\pm 120^{\circ}$ around $c^*$ (right). (c) Untwinned crystal used for the torque measurements in Figs.~\ref{aNPR:fig:calc:torque}(d) and (e). Calculations in (b) and (e) use the nominal crystal structure parameters from Ref.~\cite{synthesis}, in (e) assuming equal weight for the main grain and each of its two twins.} 
	\label{aNPR:fig:xrays}
    \end{figure}
    
    All samples used in the torque magnetometry studies were characterized by X-ray diffraction using an Oxford Diffraction Supernova diffractometer with a Mo source. \aNPR has a monoclinic crystal structure (space group $C2/c$) with room-temperature lattice parameters $a = 5.9559(2)$~\AA, $b = 10.3371(3)$~\AA, $c = 11.2069(4)$\AA, $\beta = 100.423(3)^{\circ}$, where we use the unit cell setting as defined in Ref.~\cite{synthesis}. The crystal structure is illustrated in Fig.~\ref{aNPR:fig:structure}. Pr ions occupy two nonequivalent sites, Pr(1) at Wyckoff position $4e$ at fractional coordinates $(0,0.58165(2),1/4)$ and Pr(2) $4e$ at $(0,0.24890(2),1/4)$.  

    Fig.~\ref{aNPR:fig:xrays}(a) shows the observed diffraction pattern in the $(0kl)$ reciprocal lattice plane, for a representative, untwinned single crystal. In order to prevent exposure to ambient air during the measurement of a high-statistics data set, the sample was sealed inside an evacuated borosilicate-glass capillary, and the broad ring of scattering intensity in the figure is due to scattering from the amorphous capillary walls. The sharp diffraction peaks are all indexed with the expected $C2/c$ monoclinic cell. There are no detectable streaks of diffuse scattering intensity along $l$, indicating very good structural order with no layer stacking faults, in contrast to previously reported powder samples \cite{hinatsu2006crystal, daum2021collective}, where the diffraction pattern showed evidence for frequent layer stacking faults. Structural refinement of the diffraction intensities 
    against the nominal $C2/c$ crystal structure gave consistent results to those reported earlier \cite{synthesis}. The structural refinement showed no evidence of the atomic site-mixing reported in powder samples \cite{hinatsu2006crystal, daum2021collective}.  
     
    The same synthesis batch produced a mixture of untwinned and twinned samples, which could be distinguished visually by their morphologies. Untwinned single crystals tend to grow as natural plate-like elongated hexagons as shown in Fig.~\ref{aNPR:fig:xrays}(c), with three pairs of equal length and parallel sides, with the $c^*$-axis normal to the plate and the monoclinic $a$ axis parallel to the long edges. The largest untwinned single crystals were $\sim 120\mu$m on their long edge and most samples were $\sim 10\mu$m thick. All torque measurements reported here are on untwinned single crystals. Some samples were screened with X-rays whilst sealed inside vacuumed capillaries to avoid any contact with ambient air. Other samples were covered in a thin layer of vacuum grease during the X-ray measurements, then immediately mounted on the torque platform and cooled under $^4$He exchange gas. Consistency of the torque data on several samples, with sharp transitions observed at low temperatures gave us confidence that no detectable degradation occurred to the samples during the short period of X-ray screening and/or mounting on the torque platform. 
    
    Twinned samples tend to have a more symmetric hexagonal shape as in Fig.~\ref{aNPR:fig:xrays}(f). The most common twins are rotated by $\pm120^\circ$ around a common $c^*$ axis. The diffraction pattern in the $(0kl)$ plane for a sample containing both such twins is illustrated in Fig.~\ref{aNPR:fig:xrays}(d), the reflections at $l=n\pm\frac{1}{3}$ ($n$ integer) are due to the twins, with the positions of all diffraction peaks reproduced when including the above two twins, compare Figs.~\ref{aNPR:fig:xrays}(d) with (e).         

    \begin{figure}[tb]
		\includegraphics[width=0.5\textwidth]{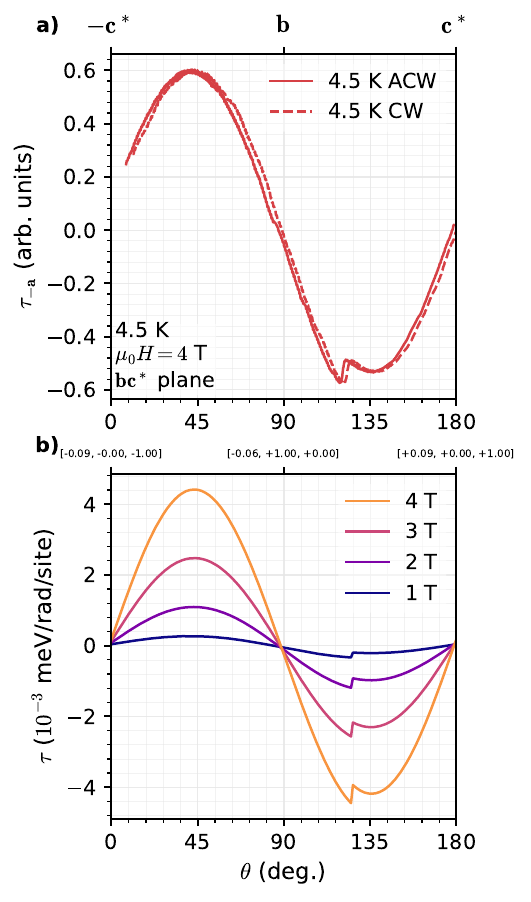}
        \centering
		\caption{(a) Torque angular dependence for field in the nominal $bc^*$ plane at 4.5~K, reproduced from Fig.~\ref{aNPR:fig:torque:bc}(b). Calculated torque angular dependence if the axis of rotation is slightly misaligned from the nominal $-a$ axis, as described in the text. Top labels indicate the field direction (in the Cartesian $xyz$ frame) for selected $\theta$ angles.}
        \label{aNPR:fig:misalignment}
	\end{figure}
    
\section{Consequences of rotation axis misalignment in the $bc^*$ plane}
\label{aNPR:sec:misalignment}
Fig.~\ref{aNPR:fig:misalignment}(b) shows the calculated torque angular dependence for the Hamiltonian parameters in Table~\ref{aNPR:calc:tab:parameters} for the case when the sample platform rotation axis $\mathbf{u}$ is slightly misaligned from the nominal $\mathbf{-a}$ axis. The consequence of this small misalignment is the appearance of a small jump in the torque curve as in the experimental data in panel (a), attributed to the field-induced switching between time-reversed magnetic domains with spontaneous magnetization along the $\pm a$ axis. The calculation assumes the rotation axis $\mathbf{u}$ is obtained from $\mathbf{-a}$ by rotation by an angle $\gamma=6^\circ$ around axis $\mathbf{\hat{n}} = \cos \delta \,\, \mathbf{\hat{c^{*}}} +
  \sin \delta\,\,\mathbf{\hat{b}}$ with $\delta=55^{\circ}$, with angles $\delta$ and $\gamma$ chosen such as to reproduce the observed $\theta$ angle of the torque jump and its magnitude relative to the full bandwidth, respectively.    

\section{Thermometer calibration in high fields}
\label{aNPR:sec:calib}
       
\begin{figure}[tb]
		\includegraphics{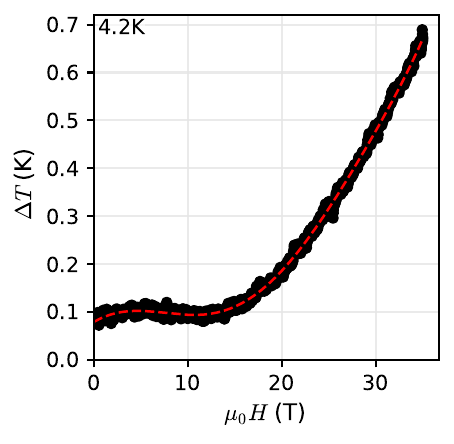}
        \centering
		\caption[Field dependence of thermometer calibration.]{Field dependence of the thermometer calibration offset measured in a bath of liquid helium at 4.2~K used to correct the temperature readings in temperature sweeps at high fields.}
		\label{aNPR:fig:nijmegen:cailbration}
	\end{figure}

For the torque measurements as a function of temperature in high magnetic fields presented in Sec.~\ref{aNPR:sec:high-field} the field-dependence of the sample thermometer has to be corrected. This was done by filling the cryostat sample volume with liquid Helium at ambient pressure to fix the temperature at 4.2~K, the offset from this was measured as a function of field and is shown in Fig.~\ref{aNPR:fig:nijmegen:cailbration}. This curve was then used to correct the measured temperatures. While there will be some temperature dependence of this correction factor, the value measured at 4.2~K is expected to be valid in the region of the phase transition close to 4.6~K.

\bibliography{aNa2PrO3_references.bib}

\end{document}